\documentclass[lettersize,journal]{IEEEtran} % Document class for IEEE Transactions

\usepackage{booktabs}

\usepackage{orcidlink}

\usepackage{glossaries}

\usepackage{amsmath, amsfonts}

\usepackage{algorithm}

\usepackage{array}

\usepackage[caption=false,font=normalsize,labelfont=sf,textfont=sf]{subfig}

\usepackage{textcomp}

\usepackage{float}

\usepackage{stfloats}

\usepackage{url}

\usepackage{verbatim}

\usepackage{graphicx}

\usepackage{cite}

\definecolor{purple}{rgb}{0.5, 0, 0.5}
\definecolor{rosepink}{rgb}{1, 0.4, 0.8}

\usepackage{wrapfig}

\usepackage{lscape}

\usepackage{rotating}

\usepackage{epstopdf}

\usepackage{tabulary}

\usepackage{enumitem}

\usepackage{algpseudocode}

\usepackage{placeins}

\usepackage{tikz}
\usepackage{pgf}
\usepackage{pgfplots}

\usetikzlibrary{patterns,shapes.arrows,calc,positioning}
\usepgfplotslibrary{groupplots,dateplot}

\pgfplotsset{compat=newest}

\graphicspath{{figure/}{figures/}{plots/}}

\newif\ifblind
\blindtrue % Uncomment for blinded version
\makeglossaries

\definecolor{mygreen}{RGB}{204,255,204}

\usepackage{afterpage} % Preliminary settings
\newacronym{BES}{BES}{battery energy storage} % \gls{BES}
\newacronym{LV}{LV}{low-voltage} % \gls{LV}
\newacronym{MPV}{MPV}{mini-photovoltaic} % \gls{MPV}
\newacronym{PV}{PV}{photovoltaic} % \gls{PV}
\newacronym{DER}{DER}{distributed energy resources} % \gls{DER}

\newacronym{VDE}{VDE}{German Association for Electrical, Electronic \& Information Technologies} % \gls{VDE}
\newacronym{NC RfG}{NC RfG}{European Network Code on Requirements for Generators} % \gls{NC RfG}
\newacronym{PCC}{PCC}{Point of Common Coupling} % \gls{PCC}

\newacronym{SoC}{SoC}{state-of-charge} % \gls{SoC}
\newacronym{RL}{RL}{Reinforcement Learning} % \gls{RL}
\newacronym{MC}{MC}{Monte Carlo} % \gls{MC}
\newacronym{VM}{VM}{Voltage Magnitude} % \gls{VM}
\newacronym{GL}{GL}{Grid Loss} % \gls{VM}
\newacronym{TL}{TL}{Transformer Loading} % \gls{TL}
\newacronym{LL}{LL}{Line Loading} % \gls{LL}
\newglossaryentry{VDE-AR-N 4105}{
	name={VDE-AR-N~4105},
	description={VDE-AR-N 4105 forms the technical basis for connecting and operating generation systems on the low-voltage grid.}
} % \gls{VDE-AR-N 4105}
\newglossaryentry{EN 50160}{
	name={EN~50160},
	description={EN 50160 Voltage characteristics in public electricity supply grids is a European standard that defines and specifies the essential characteristics of the mains voltage at the grid connection point under normal operating conditions. In Germany, the standard is valid as DIN standard DIN EN 50160.}
} % \gls{EN 50160}
\newacronym{pu}{$pu$}{per unit} % \gls{p.u} 
\newacronym{VSCs}{VSCs}{voltage source converters} % \gls{VSCs}
\newacronym{DSO}{DSO}{Distribution System Operator} % \gls{DSO}
\newacronym{RMS}{RMS}{root mean square} % \gls{RMS}
\newacronym{BIM}{BIM}{Bus Injection Model} % \gls{BIM}
\newacronym{BFM}{BFM}{Branch Flow Model} % \gls{BFM}
\newacronym{PF}{PF}{Power Factor} % \gls{PF}
\newglossaryentry{SimBench}{name={SimBench}, description={SimBench.}} % \gls{SimBench}
\newglossaryentry{MPVBench}{name={MPVBench}, description={MPVBench.}} % \gls{MPVBench}
\newglossaryentry{PandaPower}{name={PandaPower}, description={PandaPower.}} % \gls{PandaPower}
\newacronym{DWD}{DWD}{German Meteorological Service} % \gls{DWD}
\newacronym{SLP}{SLP}{standard load profiles} % \gls{SLP}
\newacronym{KPIs}{KPIs}{Key Performance Indicators} % \gls{KPIs}

\newglossaryentry{Q(V)}{
	name={$Q(V)$},
	description={Q(U) control (Q = formula symbol for the reactive power, U = formula symbol for the (grid) voltage) ensures that the inverter feeds in inductive or capacitive reactive power depending on the level of the grid voltage and thus counteracts a local voltage increase or decrease. This increases the grids' capacity to absorb additional generation systems and ensures that grid reinforcements are not required or are only required later. The active power feed-in is unaffected by the Q(U) control.}
} % \gls{Q(V)}
\newglossaryentry{Q(P)}{
	name={$Q(P)$},
	description={The Q(P) or $\cos\varphi$(P) characteristic curve adjusts the power factor ($\cos\varphi$) depending on the active power feed-in. Starting with pure active power feed-in, the cosφ is reduced above a specific power limit, and additional reactive power is drawn. The maximum reactive power occurs at maximum active power feed-in, often associated with the highest voltages at the grid connection point. This concept enables all inverters to contribute to voltage reduction, regardless of the installation location or the current voltage. In the partial load range, this results in less reactive power than with a fixed cosφ specification, but a high accumulation of reactive power can occur at full load. This reactive power reference further reduces low grid voltages.}
} % \gls{Q(P)}
\newglossaryentry{Q(P, V)}{
	name={$Q(P, V)$},
	description={The $Q(P, V)$ mode is gaining more traction~\cite{Demirok_2011} as an alternative control mode}
} % \gls{Q(P, V)}
\newglossaryentry{cosvarphi}{
	name={$\cos\varphi$},
	description={This concept allows inverters to draw reactive power regardless of their voltage. This enables PV systems near transformers to contribute to voltage reduction. However, this leads to high total reactive power, which only sometimes corresponds to the actual demand. At low grid voltages, reactive power consumption can further reduce the voltage level.}
} % \gls{cosvarphi}
\newacronym{SC}{SC}{Self-Consumption} % \gls{SC}
\newacronym{AVL}{AVL}{Automatic Voltage Limitation} % \gls{AVL}
\newacronym{PLS}{PLS}{Peak Load Shaving} % \gls{PLS}
\newacronym{DNC}{DNC}{Day-Night-Control} % \gls{DNC}
\newacronym{PR}{PR}{Penetration Rate} % \gls{PR}
\newglossaryentry{Configuration 0 (Base)}{name={Configuration 0 (Base)}, description={0}} % \gls{Configuration 0 (Base)}
\newglossaryentry{Configuration 1 (2024)}{name={Configuration 1 (2024)}, description={1}} % \gls{Configuration 1 (2024)}
\newglossaryentry{Configuration 2 (2034)}{name={Configuration 2 (2034)}, description={2}} % \gls{Configuration 2 (2034)}
\newacronym{CVR}{CVR}{Controllable Voltage Rate} % \gls{CVR}
\newacronym{TLUR}{TLUR}{Transformer Load Utilization Rate} % \gls{TLUR}
\newacronym{LLR}{LLR}{Line Loading Rate} % \gls{LLR}
\newacronym{PL}{PL}{Power Loss} % \gls{PL}
\newacronym{SITL}{SITL}{Software-In-The-Loop} % \gls{SITL}
\newacronym{AV}{AV}{Actual Value} % \gls{AV}
\newacronym{TV}{TV}{Target Value} % \gls{TV}
\newglossaryentry{15-bus}{name={15-bus},description={rural}} % \gls{15-bus}
\newglossaryentry{97-bus}{name={97-bus},description={rural}} % \gls{97-bus}
\newglossaryentry{129-bus}{name={129-bus},description={rural}} % \gls{129-bus}
\newglossaryentry{44-bus}{name={44-bus},description={semi-urban}} % \gls{44-bus}
\newglossaryentry{111-bus}{name={111-bus},description={semi-urban}} % \gls{111-bus}
\newglossaryentry{59-bus}{name={59-bus},description={urban}} % \gls{59-bus} % Acronym definitions

\begin{document}
	
	% Title, Authors, Abstract, and Keywords
	% --------------------------------------------------------------------------------
% Article Title, Authors, Abstract, and Keywords
% For IEEE Transactions on Smart Grid Submission
% --------------------------------------------------------------------------------

% Author(s): Gökhan Demirel
% Author Affiliation:
% Institute for Automation and Applied Informatics (IAI),
% Karlsruhe Institute of Technology (KIT),
% Eggenstein-Leopoldshafen, Germany
% Document root for ease of compilation
%!TEX root = ./Impact_and_Integration_of_Mini_Photovoltaic_Systems_on_Electric_Power_Distribution_Grids.tex

% --------------------------------------------------------------------------------
% I. Article Title
% The \title command allows for a "short title" option for headers.
% --------------------------------------------------------------------------------

\title{
	\footnotesize{\vspace*{-30pt} % Optional copyright notice for submitted version
		This work has been submitted to the IEEE for possible publication. \\
		Copyright may be transferred without notice, after which this version will no longer be accessible.}\\
	\vspace{25pt} % Adjusts vertical space above the main title
	{\Huge 	Impact and Integration of Mini Photovoltaic Systems on Electric Power Distribution Grids} % Main article title
}

% --------------------------------------------------------------------------------
% II. Authors
% ORCID links provided for each author.
% --------------------------------------------------------------------------------

\author{
	Gökhan~Demirel\orcidlink{0000-0002-1234-5474},
	Simon~Grafenhorst\orcidlink{0000-0002-6366-0626},
	Kevin~Förderer\orcidlink{0000-0002-9115-670X}, 
	Veit~Hagenmeyer\orcidlink{0000-0002-3572-9083},~\IEEEmembership{Member,~IEEE}
	% Acknowledgments and manuscript details
	\thanks{Manuscript received April 3, 2024; This work was supported in part by the Energy System Design (ESD) Project; in part by the Helmholtz Association’s Initiative and Networking Fund through Helmholtz AI; and in part by the Helmholtz Association Initiative and Networking Fund on the HAICORE@KIT partition. (Corresponding author: Gökhan Demirel.)}
	%	\thanks{Manuscript received April 1, 2024; This work was supported in part by the Energy System Design (ESD) Project; in part by the Helmholtz Association’s Initiative and Networking Fund through Helmholtz AI; and in part by the Helmholtz Association Initiative and Networking Fund on the HAICORE@KIT partition. Paper no. TSG-XXXXX-2024. (Corresponding author: Gökhan Demirel.)}
	\thanks{Gökhan Demirel, Simon Grafenhorst, Kevin Förderer and Veit Hagenmeyer are with the Institute of Automation and Applied Informatics IAI, Karlsruhe Institute of Technology, 76344 Eggenstein-Leopoldshafen, Germany (e-mail: goekhan.demirel@kit.edu; simon.grafenhorst@kit.edu; kevin.foerderer@kit.edu; veit.hagenmeyer@kit.edu).}
	%	\thanks{Color versions of one or more figures in this article are available at https://doi.org/10.1109/TSG.2024.XXXXXXX.}
	\thanks{Color versions of one or more figures in this article are available online.}
	%	\thanks{Digital Object Identifier (DOI): 10.1109/TSG.2024.XXXXXXX}
}
%\markboth{IEEE Transactions on Smart Grid, Vol. XX, No. XX, Month Year}
\markboth{ } 
{Demirel \MakeLowercase{\textit{et al.}}: Impact and Integration of Mini Photovoltaic Systems}
% --------------------------------------------------------------------------------
% Running Headers Configuration
% --------------------------------------------------------------------------------
\maketitle

% --------------------------------------------------------------------------------
% III. Abstract
% A concise summary of the work to be presented.
% --------------------------------------------------------------------------------

\begin{abstract}
	This work analyzes the impact of varying concentrations \gls{MPV} systems, often referred to as balcony power plants, on the stability and control of the \gls{LV} grid. 
	By local energy use and potentially reversing meter operation, we focus on how these \gls{MPV} systems transform grid dynamics and elucidate consumer participation in the energy transition. 
	We scrutinize the effects of these systems on power quality, power loss, transformer loading, and the functioning of other inverter-based voltage-regulating \gls{DER}.
	Owing to the rise in renewable output from \gls{MPV}s, the emerging bidirectional energy flow poses challenges for distribution grids abundant with \gls{DER}s.
	Our case studies, featuring sensitivity analysis and comparison of distributed and decentralized \gls{DER} control strategies, highlight that autonomous inverters are essential for providing ancillary services.
	With the growing use of \gls{BES} systems in \gls{LV} grids for these services, the need for adaptable \gls{DER} control strategies becomes increasingly evident.
\end{abstract}

% --------------------------------------------------------------------------------
% IV. Keywords
% Selected words should accurately describe the work presented.
% --------------------------------------------------------------------------------

	
	% Section I: Introduction
	\section{Introduction}
	\label{sec:introduction}
	% Author(s): Gökhan Demirel
% Author Affiliation:
% Institute for Automation and Applied Informatics (IAI),
% Karlsruhe Institute of Technology (KIT),
% Eggenstein-Leopoldshafen, Germany
%!TEX root = ./Impact_and_Integration_of_Mini_Photovoltaic_Systems_on_Electric_Power_Distribution_Grids.tex
% 001_Introduction.tex
\IEEEPARstart{B}{ALCONY POWER PLANTS}, referred to in this paper as \glsdesc{MPV} (\gls{MPV}) systems, are a pioneering innovation by Holger Laudeley in Germany and worldwide in the early 2000s~\cite{Tomik_2023_Balkonkraftwerk}.
%\IEEEPARstart{I}{n} the early 2000s, Holger Laudeley pioneered the promotion and advocacy of \glsdesc{MPV} (\gls{MPV}) systems in Germany and worldwide~\cite{Tomik_2023_Balkonkraftwerk}.
Since then, their prevalence has grown exponentially. 
His vision of easily installed, portable mini balcony power units has materialized into a significant trend in the renewable energy sector. 
Taking Germany as an example, an estimated 250,000 units will be in operation in 2024 alone, contributing to an energy-saving potential of $100$~MWh.
Furthermore, \gls{PV} systems constitute approximately $45$ percent of the total installed renewable energy capacity of $150$~GWp~\cite{BMWK_2023}.
To further promote the adoption of these systems, the \gls{VDE} has proposed amendments, including raising the maximum power to $800$~Wp and implementing a regulation allowing meters to run backward within this limit.
These proposed changes align with the \gls{NC RfG}), as described in~\cite{EU2016_631} and \cite{VDE_2023}, and they are expected to boost the penetration rate of plug-and-play MPV systems.
Several European countries, including Belgium, Cyprus, Denmark, Italy, and the Netherlands, have adopted net metering as a standard practice~\cite{Poullikkas_2013, Gautier_2018}. 
Moreover, 44 US states also use this approach~\cite{dsire_2023}.
This widespread use highlights the global importance of the German households \gls{VDE} changes.
With solar panels achieving grid parity for many households~\cite{Bergner_2022}, the self-consumption of locally generated \gls{PV} electricity has become more economically viable than relying on feed-in tariffs into the electrical grid~\cite{BNetzA_2023}.
Using \glsdesc{BES} (\gls{BES}) systems further amplifies this advantage~\cite{Geth_2010}.
As a result, the economic appeal and straightforward integration of local balcony power plants encourage increased household investments.
%-Ziel: Kurze Erwähnung von Herausforderungen/Bedenken bei MPV-Systemen.
%-Kontext_ Integration in bestehende Energieinfrastruktur.
%-Nutzen: Abgerundeter Blick auf das Thema.
%-Mehrwert: Aufzeigen von Forschungs- oder Politikbereichen für die Zukunft. => sucess
Prospects for widespread adoption of \gls{MPV} systems are looking positive.
Active research and progressive regulatory changes are addressing issues such as grid stability, safety concerns, and end-of-life panel management, continually enhancing the viability and sustainability of these systems~\cite{BMWK_2023,EU2016_631,VDE_2023,Bergner_2022,BNetzA_2023}.

%\subsection{Research Questions}
This paper explores the complex dynamics of \gls{LV} grid operation with an increasing penetration and integration of \gls{MPV}s and \glsdesc{DER} (\gls{DER}s), considering their interactions with other \gls{DER}s.
As the global community steers toward a sustainable future, comprehending the consequences of assimilating innovative energy sources gains significance. 
%Such investigations represent a novel area of interest in the energy domain.
Expectations are that gaining insights into these dynamics will illuminate critical aspects of operational stability, safety, and efficiency in power systems, especially at the \gls{LV} grid level.
These insights are currently scarce as \gls{MPV}s' are a relatively new kind of \gls{DER}, which only recently started to gain traction and with seemingly little impact.
By addressing the following main research questions, we aim to illuminate and outline the outcomes of such an integration. 

The questions serve as guiding posts for the study, intending to shed light on potential impacts while offering fundamental strategies for optimization and control of power systems amidst growing \gls{MPV} and \gls{DER} integration:\\
\begin{enumerate}
	\item What are the effects of increasing \gls{MPV} and \gls{DER} integration on grid stability and reverse power flow, considering the radial topology of \gls{LV} Grids?
	\item How does \gls{MPV} integration at the \gls{PCC} influence voltage rise in an \gls{LV} Grid, a phenomenon crucial for grid operational efficiency and safety?
	\item What inverter control strategies can be employed for optimized self-consumption and effective grid control in the face of growing \gls{MPV} proliferation?
	\item How might (de facto) voltage regulatory frameworks be updated to deal with grid constraints related to the increasing integration of \gls{MPV} and \gls{DER} and its challenges?
\end{enumerate}
This research aims to contribute to the renewable energy sector and electrical grid management concerning \gls{MPV} integration. The findings provide guidances for strategies for leveraging these uncertain \gls{MPV}s and \gls{DER}s due to weather-dependent power generation while ensuring grid stability and safety.

%\subsection{Contributions}
In summary, the main contributions of this paper are as follows:
\begin{itemize}
\item  We present an open-source evaluation environment~\footnote{\footnotesize The code used for this research will be published with this paper on our GitHub repository online: %https://github.com/demirelg/powergrid-vde-ieee-control
\url{} (\textit{blinded for review})} developed with Python, implementing the \gls{VDE}-AR-N 4105 guidelines, to facilitate the decentralized provision of system service powers (active and reactive) at \gls{DER} inverter terminals.
\item We research challenges associated with \gls{MPV}s, evaluate rule-based and time-of-day \gls{BES} control strategies, and different \gls{BES} sizing to mitigate congestion and enhance operator profit across various \gls{DER} penetration configurations in \gls{LV} grids.
\item We conduct real-world time series data analysis and extensive simulations across various \gls{LV} grid topologies ranging from small rural (\gls{15-bus}) to large urban configurations (\gls{59-bus}). 
We also provide the dataset \gls{MPVBench} of real-time \gls{MPV} data for the energy community with associated metadata\footnote{\footnotesize The full benchmark dataset can be downloaded from our GitHub repository online: % https://github.com/demirelg/MPVBench
\url{} (\textit{blinded for review}) }, which contains detailed information on the hardware components.
% utilized, such as solar panels, inverters, and meters.
\item We set up our environment for \gls{RL} and AI control compatibility, enabling adaptable electrical grid control and analysis, and allowing AI solution comparison and integration in different grids. %electric power distribution grids.
\end{itemize}
To address the active voltage control problem, we compare various decentralized strategies, aiming to suggest potential directions for future research. 
Our environment, designed for flexibility, allows for swift extension with additional grid topologies and data, and is using \gls{PandaPower}~\cite{Thurner_2018} and \gls{SimBench}~\cite{Meinecke_2020}.
%\subsection{Structure}

The structure of this paper is as follows: Section~\ref{sec:RelatedWorks} addresses challenges and solutions in active voltage regulation. 
Section~\ref{sec:background} outlines the regulatory framework for the European electrical grid system. 
In Section~\ref{sec:Problem Formulation}, we introduce the approach focussing on the influence of \gls{MPV}s on voltage regulation in \gls{LV} grids.
Section~\ref{sec:PowerSystemEnvironment} delves into \gls{BES} uncertainty modeling and \gls{MPV} integration.
Section~\ref{sec:InverterControlStrategiesinVDE-AR-N4105} explores Grid Code's local \gls{DER} control modes, followed by diverse \gls{DER} control strategies.
Section~\ref{sec:CaseStudies} presents the case studies, including experimental settings, data and grid topology description, performance metrics, and a description of our grid control evaluation environment.
Finally, Section~\ref{sec:Conclusion} outlines the main conclusions and future research directions.
	
	% Include figure (example: fig01)
	% Author(s): Gökhan Demirel
% Author Affiliation:
% Institute for Automation and Applied Informatics (IAI),
% Karlsruhe Institute of Technology (KIT),
% Eggenstein-Leopoldshafen, Germany
%!TEX root = ./Impact_and_Integration_of_Mini_Photovoltaic_Systems_on_Electric_Power_Distribution_Grids.tex
% fig01.tex
\begin{figure}
	\centering
	{
	\fontsize{10pt}{9pt}\selectfont% or whatever fontsize you like
	\def\svgwidth{3.53in}
	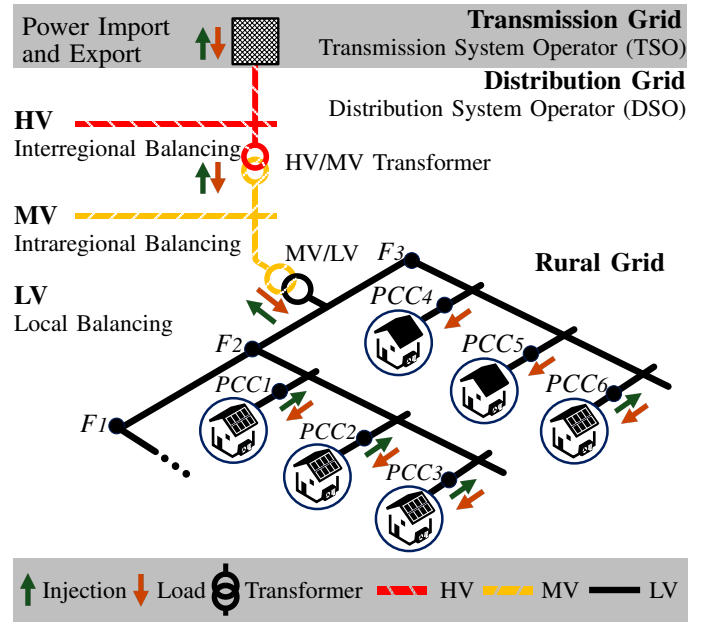
	} 
%	\vspace*{-5mm}
	\caption{Illustration of the European Power System showing interactions and connections via transformers between the transmission grid, inclusive of extra high voltage, the distribution grid, comprising high-voltage (indicated in red), medium-voltage (shown in yellow), and \gls{LV} (represented in black). Power absorption is depicted in red arrows, while power injection into the higher and lower levels is indicated with green arrows.
	}
	\label{fig:01figEuropeanPowerSystem}
\end{figure}
	
	% Section II: Related Work
	\section{Related Work}
	\label{sec:RelatedWorks}
	% Author(s): Gökhan Demirel
% Author Affiliation:
% Institute for Automation and Applied Informatics (IAI),
% Karlsruhe Institute of Technology (KIT),
% Eggenstein-Leopoldshafen, Germany
%!TEX root = ./Impact_and_Integration_of_Mini_Photovoltaic_Systems_on_Electric_Power_Distribution_Grids.tex
% 002SectionRelatedWorks.tex
As the twilight of traditional grids approaches, the voltage control problem has been gaining momentum due to the increased integration of distributed resources like rooftop \gls{PV}s~\cite{Omran_2011}.
Despite a declining growth rate, Germany is among the leading countries in installed \gls{PV} capacity per capita, catalyzing further research endeavors for future high \gls{PV} penetration configurations~\cite{stetzthomas_2015}.
Voltage rise in distribution systems with small generators has been a focus for over two decades~\cite{Masters_2002}. 
Decentralized power generation in \gls{LV} grids can cause reversed load flow and overvoltages, which can be mitigated by the reactive power consumption of \gls{VSCs}~\cite{Kerber_2009}. 
Cloud-induced transients in \gls{PV} power can cause voltage flicker or excessive operation of voltage regulating equipment~\cite{Ari_2011}. 
However, high \gls{PV} penetration does not adversely impact grid voltage when distributed \gls{PV} resources do not exceed an average of 2.5 kW per household on a typical distribution grid~\cite{Tonkoski_2012}.
While they focus only on \gls{PV}s, we analyze \gls{MPV}s in the \gls{LV} grid to highlight their impact on grid dynamics and challenges in grids with various penetrations of \gls{DER}s.
Active and reactive power control strategies can effectively mitigate voltage rise and expand \gls{LV} grid capacity~\cite{Stetz_2013}. 
Local \gls{PV} storage and voltage control strategies also enhance \gls{PV} grid integration~\cite{von_Appen_2014}, while autonomous voltage control strategies can defer grid reinforcement for economic benefits~\cite{Stetz_2014}. 
Both~\cite{Kabir_2014} and~\cite{Zeraati_2018} propose coordinated control strategies using \gls{PV} and \gls{BES} systems to regulate voltage in \gls{LV} grids with high \gls{PV} penetrations.
\cite{Kabir_2014} shows that reactive \gls{PV} inverters in combination with \gls{BES} systems 
can control voltage rise and drop issues with a droop-based method.
Conversely, \cite{Zeraati_2018} targets voltage rise during peak \gls{PV} generation and voltage drop during peak load. 
The voltage profile of a distribution feeder under high \gls{PV} penetration, as per Brazilian regulations, is analyzed in~\cite{Vargas_2018}. 
A voltage sensitivity analysis similar to~\cite{Demirok_2011} is necessary to determine optimal configuration parameters like critical voltage values. 
Furthermore, \cite{Matkar_2017} observed an increase in voltage rise with higher \gls{DER} penetration levels, attributing this to the line impedance. 
Notably, the most distant \gls{DER} integration may cause the highest voltage rise~\cite{Singh_2020}. 
Moreover, the $X/R$ ratio of a \gls{LV} distribution grid line is relatively low, so 
neither the active ($RP$) nor the reactive terms ($XQ$) are negligible, and these terms can influence the voltage level at the \gls{PCC}~\cite{Masters_2002}.
In~\cite{Biel_2018} introduces a single-phase inverter with reactive power control, as also highlighted in~\cite{Wang_2018}, which presents a hybrid modulation method for single-phase DC-AC conversion within new grid codes for ancillary services.
%The work by~\cite{Biel_2018} presents the regulation of active and reactive power in single-phase \gls{PV} inverters.
%Some new grid codes require single-phase inverters to use reactive power for providing ancillary services~\cite{Wang_2018}, that converts $DC$ source voltage into single-phase $AC$ output voltage at a desired voltage.
Inspired by these related contributions, this paper aims to investigate the impact of \gls{MPV} integration on the \gls{LV} grid, including active voltage control of \gls{DER}s and maintaining grid resilience.

	% Section III: Background
	\section{Background}
	\label{sec:background}
	% Author(s): Gökhan Demirel
% Author Affiliation:
% Institute for Automation and Applied Informatics (IAI),
% Karlsruhe Institute of Technology (KIT),
% Eggenstein-Leopoldshafen, Germany
%!TEX root = ./Impact_and_Integration_of_Mini_Photovoltaic_Systems_on_Electric_Power_Distribution_Grids.tex
% 003SectionBackground.tex
\subsection{Regulatory Framework} 
\label{sec:RegulatoryFrameworkandGridCodes}
In Germany, the regulations and technical guidelines for static voltage stabilization comply with the voltage quality limits specified in the standards~\gls{EN 50160} and the \gls{VDE} application guideline~\gls{VDE-AR-N 4105}, which applies to the grid operation of \gls{DER}s on the \gls{LV} grid. 
The \gls{LV} grid operates at a voltage level of $V_{\text{LV}}$ $\leq$ $1$~$kV$ and has a nominal grid voltage ($V_{\text{nom}} = 230~V$). 
The \gls{DSO} allows \gls{DER}s to operate within \gls{EN 50160}'s defined limits of $0.85$ to $1.10$ volts \gls{pu} from the nominal voltage during undisturbed grid operation; otherwise, they automatically disconnect from the grid.
\gls{EN 50160} defines that the \gls{RMS} of the 10-minute average voltage shall ensure that the voltage range remains within $V_{\text{nom}}$ $\pm 10~\%$ for $95~\%$ of each week~\cite{DIN_EN50160_2020}. \gls{DSO}s adhere to this standard and ensure that voltage drops between lines and transformers remain within this range.
\gls{VDE-AR-N 4105} mainly defines a regulation that allows a maximum voltage rise of $3$~$\%$ at each \gls{PCC} due to \gls{DER} installations in \gls{LV} grids.
\gls{DSO}s prescribe methods for the feed-in of reactive power for static voltage stabilization at the terminals of the inverter-based \gls{DER}~\cite{VDE_2018}.
%, as detailed in Table~\ref{tables:DER Types} of Appendix~\ref{appendix:ReactivePowerRequirements}.
\subsection{System Architecture}
\label{sec:SystemArchitecture}
Fig.~\ref{fig:01figEuropeanPowerSystem} visually represents relationships and the transformation of the European Power System that enables the grid to supply electricity and ancillary services to its users and the primary grid, as depicted by the bidirectional power flows. 
Administrative units, typically the \gls{DSO}, monitor and operate these \gls{PV} systems via communication channels.
Driven by recent changes in European grid code, the fast and seamless integration of balcony power plants into existing systems has become possible through plug-and-play technology using household sockets. 
In Section \ref{sec:ExperimentalSettings}, we will consider different \gls{MPV} parameterizations in the \gls{LV} grid. We then perform a sensitivity analysis to evaluate the influence of \gls{MPV} parameters on the electrical grid.

	% Author(s): Goekhan Demirel
% Institution(s): Institute for Automation and Applied Informatics (IAI)
%!TEX root = ./Impact_and_Integration_of_Mini_Photovoltaic_Systems_on_Electric_Power_Distribution_Grids.tex
% fig02.tex
\begin{figure}[t]
	\centering
%	\resizebox{9.0cm}{5.0cm}{
%		{
%		\fontsize{31pt}{50pt}\selectfont% or whatever fontsize you like
%		\def\svgwidth{15.5in} 
%		\input{figures/pdf/02figEquivalentPowerCircuit_v2.pdf_tex}
%		}
%			}
	\def\svgwidth{1.00\columnwidth}
	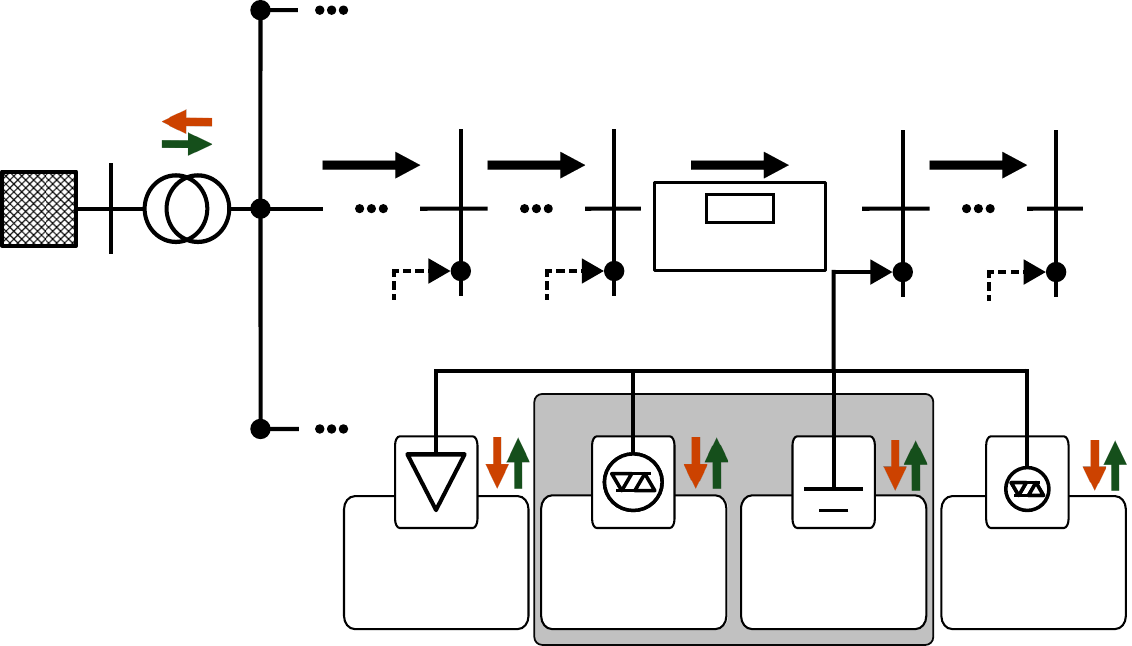
%	\vspace*{-5mm}
	\caption{Equivalent circuit in a radial electrical distribution system including Loads, \gls{PV}s, \gls{BES}s, and \gls{MPV}s.}
\label{fig:02figEquivalentPowerCircuit}
\end{figure}
\FloatBarrier % Include figure (example: fig02)
	
	% Problem Formulation Sections IV.1 and IV.2
	% Author(s): Gökhan Demirel
% Author Affiliation:
% Institute for Automation and Applied Informatics (IAI),
% Karlsruhe Institute of Technology (KIT),
% Eggenstein-Leopoldshafen, Germany
 %!TEX root = ./Impact_and_Integration_of_Mini_Photovoltaic_Systems_on_Electric_Power_Distribution_Grids.tex
 % 004Section01PROBLEMFORMULATION.tex
 \subsection{Active Voltage Control Problem}
 \label{sec:Problem Formulation}
 \label{subsec:Active Voltage Control Problem}
We consider the \gls{LV} distribution grid (represented in black) in Fig.~\ref{fig:01figEuropeanPowerSystem} with arbitrary topologies as a graph $\mathcal{G} = (\mathcal{V},\mathcal{E})$ with a set of nodes (buses) $\mathcal{V} = \{ 0, 1, \ldots, n \}$ with $n \in \mathbb{N}^{+}$. Each element of the set of edges (or lines) $\mathcal{E} = \{ 1, \ldots, n\}$ is a pair of nodes $\mathcal{E} \subseteq \mathcal{V}~\times~\mathcal{V}$. 
A typical European \gls{LV} grid is characterized by a three-phase radial topology. 
In the present paper, we simplifiy the European \gls{LV} grid to a single-phase system, representing it as a tree graph of $n+1$ buses in Fig.~\ref{fig:02figEquivalentPowerCircuit}.
The bus indexed with~$0$ denotes the tree's root, representing the primary slack or substation bus connected to the superior electrical grid level. 
It regulates and balances the active and reactive power within the \gls{LV} grid.
This bus $V_{0}$ typically maintains a fixed reference voltage.
Let $\mathcal{E}_{\pi(j)} \subseteq \mathcal{V}$ be the set of all nodes that are parents to the node $j$ and $\mathcal{E}_{\sigma(j)} \subseteq \mathcal{V}$ the set of all nodes that are children of the node $j$.
The complex power flowing from bus $i$ to bus $j$ is denoted by  $S_{ij} = P_{ij} +\textbf{j}Q_{ij}$, and can be decomposed into its active ($P_{ij}$) and reactive ($Q_{ij}$) components.
At each bus $i \in \mathcal{V}$, $V_{i}$ and $\theta_{i}$ denote the magnitude and phase angle of the complex voltage.
Let $s_{j} = p_{j} +\textbf{j}q_{j}$ be the net complex power injection at bus $j$. In this context, Tellegen's theorem, an offshoot of Kirchhoff's laws, elucidates the interrelationships between generated power, load flow, net power injection, and power transmission to and from bus $j$ and its adjacent buses:
\begin{subequations}
\begin{flalign}
	p_{j} & = p_{j}^{\text{G}} - p_{j}^{\text{C}} =& p_{j}^{\text{MPV}} + p_{j}^{\text{PV}} + p_{j}^{\text{BES}} - p_{j}^{\text{Load}} & = p_{j}^{\text{T}} \ \ \ \forall j \in \mathcal{V}, \label{subeq:2a} \\
	q_{j} & = q_{j}^{\text{G}} - q_{j}^{\text{C}} =& q_{j}^{\text{MPV}} + q_{j}^{\text{PV}} + q_{j}^{\text{BES}} - q_{j}^{\text{Load}} & = q_{j}^{\text{T}} \ \ \ \forall j \in \mathcal{V}, \label{subeq:2b} 
\end{flalign}
\end{subequations}
where terms $p_j$ and $q_j$ denote local generation by \gls{MPV} and \gls{PV} as well as \gls{BES} discharging at node $j$ minus local loads and \gls{BES} charging, which is included in $p_j^{BES}$ and $q_j^{BES}$.
\gls{PV} and \gls{BES} systems are inverter-based \gls{DER} units, while MPV systems only supply active power. 
The active and reactive power of \gls{BES} systems, symbolized as $p_{j}^{\text{BES}}$ and $q_{j}^{\text{BES}}$, can be exported to or imported from the grid.  
Energy exchange in the \gls{BES} follows the relationship~\(p_{j}^{\text{BES}} = p_{j}^{\text{BES},\text{dis}} + p_{j}^{\text{BES},\text{cha}}\), in which the term \(p_{j}^{\text{BES},\text{dis}} \geq 0\) denotes the discharging power (energy injection into the electrical grid) and \(p_{j}^{\text{BES},\text{cha}} \leq 0\) the charging power (energy absorption from the electrical grid).
This sign convention is used for \gls{DER}s and clarifies that we consider energy provision as positive and energy consumption as negative.
For loads, a positive value signifies energy consumption from the electrical grid. 
The total \gls{DER} reactive power, denoted by ($q_{j}^{\text{DER}} = + q_{j}^{\text{PV}} + q_{j}^{\text{BES}}$), has a dual influence on voltage. 
Similarly, the active power is represented by $p_{j}^{\text{DER}}$.
When the reactive power is in the under-excited state, which is characterized by the absorption of reactive power ($- q_{j}^{\text{DER}}$), there is a drop in voltage due to inductive consumption.
On the other hand, the injection of reactive power ($+ q_{j}^{\text{DER}}$) in the overexcited state leads to a rise in voltage due to the capacitive consumption of the \gls{DER} inverters.
Contrarily, loads absorb active and reactive power, captured by $p_{j}^{\text{Load}}$ and $q_{j}^{\text{Load}}$. 
In cases where the bus $j$ has no load, we assume them to be zero.
Similarly, we directly zero the respective values in the absence of \gls{MPV}, \gls{PV}, or \gls{BES} generation at bus $j$.
Power flow problems describe the full non-linear AC power flow Eqs. and are essential for grid calculation and control. 
Two fundamental formulations play a role here and are equivalent to each other: The power flow Eqs., formally known as the \gls{BIM}, is equivalent to the \gls{BFM}~\cite{Subhonmesh_2012}. 
Building on these fundamentals, the relaxed \gls{BFM} Eqs. are as follows~\cite{Farivar_2013}:
\begin{subequations} 
	\begin{align}
		p_{j} &=  
		\sum_{i\in \mathcal{E}_{\pi(j)} } (r_{ij}l_{ij}-P_{ij}) + \sum_{k\in \mathcal{E}_{\sigma(j)} } P_{jk},  &\forall j \in \mathcal{V}, \label{subeq:1a} \\
		q_{j} &=  
		\sum_{i\in \mathcal{E}_{\pi(j)} } (x_{ij}l_{ij}-Q_{ij}) + \sum_{k\in \mathcal{E}_{\sigma(j)} } Q_{jk},  &\forall j \in \mathcal{V}, \label{subeq:1b} \\
		V_{j}^{2} &= V_{i}^{2} - 2(r_{ij}P_{ij}+x_{ij}Q_{ij}) + (r_{ij}^{2}+x_{ij}^{2})l_{ij},  &\forall (i,j) \in \mathcal{E},
		\label{subeq:1c} \\
		l_{ij}^{2} &= \frac{P_{ij}^{2}+Q_{ij}^{2}}{V_{i}^{2}},  &\forall (i,j) \in \mathcal{E},  
		\label{subeq:1d} 
	\end{align}
\end{subequations}
where $V_{j}$ is the voltage at the \gls{PCC} node $j$, and $z_{ij}=r_{ij}+\textbf{j}x_{ij}$ represents the impedance $z_{ij}$ described by the resistance $r_{ij}$ and reactance $x_{ij}$ on the line connecting buses $i$ and $j$. 
Based on Kirchhoff's current and voltage laws, the first two Eqs. (\ref{subeq:1c}) and (\ref{subeq:1d}) balance active and reactive power. 
Ohm's law is the base for Eq. (\ref{subeq:1c}), and Eq. (\ref{subeq:1d}) that defines $l_{ij}^{2}$ as the squared magnitude of the complex branch current on the line from bus $i$ to $j$.

	% Author(s): Gökhan Demirel
% Author Affiliation:
% Institute for Automation and Applied Informatics (IAI),
% Karlsruhe Institute of Technology (KIT),
% Eggenstein-Leopoldshafen, Germany
%!TEX root = ./Impact_and_Integration_of_Mini_Photovoltaic_Systems_on_Electric_Power_Distribution_Grids.tex
% 004Section02PROBLEMFORMULATION.tex
 \subsection{Regulatory Control Constraints}
\label{subsec:Regulatory Control Constraints}
From the regulatory perspective, among the five local components contributing to $p_{j}+\textbf{j}q_{j}$, four --- namely $p_{j}^{\text{Load}}$, $p_{j}^{\text{MPV}}$, $p_{j}^{\text{DER}}$, and $q_{j}^{\text{Load}}$ --- are given uncontrollable quantities. 
In contrast, inverter-based \gls{DER} control algorithms actively regulate the dynamics of reactive power flow \(q_{j}^{\text{DER}}\), crucial for maintaining the voltage stability of the power grid.
Reactive power transmission is subject to restrictions, resulting in voltage regulation by \gls{DER}s in the LV grid.
The reactive power control indirectly influences the active power $p_{j}^{\text{DER}}$, with the separate inverters apparent powers for \gls{PV} (\(s_{j}^{\text{PV}}\)) and \gls{BES} (\(s_{j}^{\text{BES}}\)) systems determining the operating constraints:
\begin{align} \label{equation:3}
\left|q_{j}^{\text{DER}}\right| \leq \sqrt{(s_{j}^{\text{PV}})^2 - (p_{j}^{\text{PV}})^2} + \sqrt{(s_{j}^{\text{BES}})^2 - (p_{j}^{\text{BES}})^2}
\end{align}
The voltage drop $\Delta V_{ij} = V_{i} - V_{j}$ across a distribution line connecting nodes $i$  and $j$ can be approximated using the relationship~\cite{Kerber_2009}:
\begin{align} \label{equation:4}
\Delta V_{ij}  = \frac{r_{ij}  \left( p_{j}^{\text{Load}} - p_{j}^{\text{DER}} \right) +x_{ij}\left( q_{j}^{\text{Load}} -q_{j}^{\text{DER}}\right)}{V_{j}}
\end{align}
where $V_{i}$ is the parent busbar's voltage, which serves as a reference value. 
The terms \( r_{ij} \) and \( x_{ij} \) denote the resistance and reactance, respectively, between buses \( i \) and \( j \) as depicted in Fig.~\ref{fig:02figEquivalentPowerCircuit}.
Predominantly, such voltage deviations are contingent on the distribution line impedance combined with the dynamics of active power feed-in~\cite{Akinyemi_2022}.
Given Eq. (\ref{equation:4}), we describe the active power loss of the distribution line $j$ connecting buses $i$ and $j$ as follows~\cite{Baran_1989_sizing,Turitsyn_2011}:
\begin{align} \label{equation:5}
P_{j}^{\text{loss}} = \frac{\left( p_{j}^{\text{Load}} - p_{j}^{\text{DER}} \right)^{2} + 
	\left( q_{j}^{\text{Load}} -q_{j}^{\text{DER}}\right)^{2} }{ V_{i}^{2} } 
\cdot r_{ij} 
\end{align}
Two pivotal objectives take center stage in the global characteristics of electrical distribution grids.
The primary goal is to maintain the voltage regulation within a certain range to ensure safe and optimal operation.
We can express these voltage range constraints:
\begin{align} \label{equation:6}
V_{0}-\epsilon_{v} \leq V_{j} \leq V_{0}+\epsilon_{v}  \quad \forall j \in \mathcal{V} \quad \text{\gls{pu}}
\end{align}
where $\epsilon_{v}$ is approximately $0.05$~\gls{pu}, as described in Section~\ref{sec:RegulatoryFrameworkandGridCodes}.
During nocturnal high-load intervals, the end consumer voltage can drop below $0.95$~\gls{pu}, which represents an increased demand for electricity~\cite{Agalgaonkar_2014}; conversely, in situations with significant feed-in from $p_j^{\text{DER}}$, the electricity export process leads to a reverse current flow that causes $V_{j}$ to voltage rise over its nominal range~\cite{Masters_2002}.
This dichotomy underscores the inherent challenges of balancing power generation with consumption patterns. 
The total active power losses in the \gls{LV} grid, which corresponds to the aggregated line losses from Eq.~(\ref{equation:5}), are given by~\cite{Baran_1989_sizing} and~\cite{Baran_1989_loss}:
%by~\cite{Baran_1989_sizing, Baran_1989_loss}:
\begin{align}
\label{equation:7}
\mathcal{P}^{\text{loss}} = \sum_{i=0}^{n-1} r_{ij} \frac{(P_{ij}^{2}+Q_{ij}^2)}{V_{i}^{2}} = \sum_{i=0}^{n-1} r_{ii} l_{ij}^{2} \quad \text{\gls{pu}}
\end{align}
where \(\mathcal{P}^{\text{loss}}\) denotes the total active power loss in the grid. 
Minimizing this loss is a secondary yet paramount objective in the electrical distribution grid.
	
	% Section V: Electrical System Environment
	\section{Electrical System Environment}
	\label{sec:PowerSystemEnvironment}
	% Author(s): Gökhan Demirel
% Author Affiliation:
% Institute for Automation and Applied Informatics (IAI),
% Karlsruhe Institute of Technology (KIT),
% Eggenstein-Leopoldshafen, Germany
%!TEX root = ./Impact_and_Integration_of_Mini_Photovoltaic_Systems_on_Electric_Power_Distribution_Grids.tex 
% 005Section01PowerSystemEnvironment.tex
In this section, we deal with the complexity of the electrical grid, focusing on the \gls{BES} and inverter models for \gls{DER}s, as the environment changes over time. 
The \gls{BES} subsection \ref{sec:BESSystemModelwithUncertainty} describes a grid-connected and economically operating storage system and focuses on the essential parts of the stochastic uncertainties and the energy balance Eqs. of \gls{BES}. 
%Subsequently, we then define and integrate the inverter model \gls{DER} into the environment. 
%This \gls{DER} inverter model includes the efficiency of the inverter, the communication dynamics, and inherent uncertainties.
We denote our equations with a subscript \( t \) to indicate time-dependent variables.
	% Author(s): Gökhan Demirel
% Author Affiliation:
% Institute for Automation and Applied Informatics (IAI),
% Karlsruhe Institute of Technology (KIT),
% Eggenstein-Leopoldshafen, Germany
%!TEX root = ./Impact_and_Integration_of_Mini_Photovoltaic_Systems_on_Electric_Power_Distribution_Grids.tex
% 005Section02BESSystemModelwithUncertainty.tex
\subsection{\gls{BES} System Model with Uncertainty}
\label{sec:BESSystemModelwithUncertainty}
A typical \gls{BES} system is designed with two main goals: to be economically efficient and to operate within grid constraints. 
This design aims to optimize self-consumption for the system owners.
The \gls{BES} Model with the energy storage formula is expressed in Eq. (\ref{equation:BES-1}), where $p_{t}^{\text{dis}}$ is the discharging power and $p_{t}^{\text{cha}}$ is the charging power of the \gls{BES}, respectively; and $E_{t}$ is the energy stored in the battery at time $t$;  $\Delta t$ represents the time difference between the previous and the current time step.
In addition, the $E^{\max}$ represents the \gls{BES}'s maximum energy capacity, while $E^{\min}=0$ denotes its minimum and indicates that the \gls{BES} cannot store negative energy.
\begin{align}\label{equation:BES-1}
	E_{t} = E_{t-1} - (\eta^{\text{cha}} \cdot p_{t}^{\text{cha}} + \frac{p_{t}^{\text{dis}}}{\eta^{\text{dis}}}) \cdot \Delta t - \eta^{\text{self}} \cdot E^{\max} %\quad E_{t} \in \{0,E^{\max}\}
\end{align}
where the coefficients $\eta_{cha}$ and $\eta_{dis}$ are the discharge and charge efficiencies, and a relative self-discharge rate $\eta^{self}$ is subject to stochastic uncertainty parameters and considered per time step.
This trio coefficients $\eta^{cha}$, $\eta^{dis}$ and $\eta^{self}$ are treated as random variables, each following a normal distribution:
\begin{subequations}
	\label{subequation:BES-2}
	\begin{align}
		\eta^{\text{cha}} &\sim \mathcal{N}(\mu_c, \sigma_c^2)
		\label{subequation:BES-2a} \\
		\eta^{\text{dis}} &\sim \mathcal{N}(\mu_d, \sigma_d^2)
		\label{subequation:BES-2b} \\
		\eta^{\text{self}} &\sim \mathcal{N}(\mu_s, \sigma_s^2)
		\label{subequation:BES-2c} 
	\end{align}
\end{subequations}
where \(\mu_c\), \(\mu_d\), and \(\mu_s\) are the respective mean values, and \(\sigma_c\), \(\sigma_d\), and \(\sigma_s\) denote the corresponding standard deviations. 
These pairs of variables each define a unique normal distribution, assumed based on the law of large numbers theorem.
This approach takes into account the nonlinear and uncertain properties of the \gls{BES} model.
It contains typical system noise and measurement errors that are based on the standard deviation of $1$~\%, which is common in \gls{PCC} voltage metering~\cite{VDE_2018, IEEE_2018}.
This deviation represents a $\sigma$ of $0.01$~$\mu$.
It encapsulates both the inherent variability in \gls{BES} modeling parameters and the probabilistic fluctuations observed in operation, such as the \gls{SoC} and the reachable output power of a \gls{BES}.
Randomly generated normally distributed errors, as represented in the Eqs. (\ref{subequation:BES-2a} - \ref{subequation:BES-2c}), are estimated with the set of standard deviations $\sigma =\ [ \sigma_{c}, \sigma_{d},\sigma_{s} ] =\left\{ \ 0.01~\mu_{c}, 0.01~\mu_{d}, 0.01~\mu_{s} \ \right\}$, fostering a more robust and realistic simulation of system behavior.
The \gls{BES} capacity and efficiency decline over time due to aging effects and a significant rise in internal resistance~\cite{Naumann_2015}.
Furthermore, the self-discharge rate of BES is typically 6--7~\% per month~\cite{2015_Schmidt, Naumann_2015}.

The \gls{BES} uncertainty parameters, which can be changed or neglected in the configuration file, enable high adaptability when tuning the electrical system model to different use case configurations and an optional storage scaling factor configured to 1.0 by default. 
In electrical grid system simulation, \gls{MC} methods use stochastic parameters to ensure accurate random sampling~\cite{monte_1949}.
In \gls{RL}, agents perform control actions while considering uncertainties in the environment~\cite{Sutton_2018}.
For the operational safety of the electrical grid, maintaining the maximum power limits during charging and discharging is crucial for the operating dynamics of the \gls{BES}.
These operational \gls{BES} constraints are given in Eqs. (\ref{subequation:BES-3}) and (\ref{subequation:BES-4}), as well as the bus feed-in and feed-out power limitation in Eq. (\ref{subequation:BES-5}).
\begin{align}
	0 \leq        	  -&p_{t}^{\text{cha}} \leq p^{\text{cha}-\max}  \label{subequation:BES-3} \\
	0 \leq 		      \phantom{-}&p_{t}^{\text{dis}} \leq p^{\text{dis}-\max}  \label{subequation:BES-4} \\
	p_{t}^{\min} \leq \phantom{-}&p_{t}     		  \leq p_{t}^{\max}  		\label{subequation:BES-5}
\end{align}
where $p^{\text{cha}-\max}$ and $p^{\text{dis}-\max}$ represent the maximum active power during charging and discharging, and $p_t^{\min}$ and $p_t^{\max}$ denote the upper and lower boundaries of bus injection.
The bus feed depends on the load and \gls{DER} generation and results from the input time series for each time step.
Eq. (\ref{subequation:BES-6}) ensures independent charging and discharging. 
Specifically, $p_{t}^{\text{cha}}$ and $p_{t}^{\text{dis}}$ are mutually exclude at any given time step $t$, as Eq. (\ref{subequation:BES-6}) shows mathematically: 
\begin{align} \label{subequation:BES-6}
	p_{t}^{\text{cha}} \cdot p_{t}^{\text{dis}} = 0 
\end{align}
%Equation (\ref{subequation:BES-7}) shows a relationship between the model and the minimum and maximum limits of the stored energy:
%\begin{align} \label{subequation:BES-7}
%	E^{\min} \leq &E_{t} \leq E^{\max}
%\end{align}
%where $E^{\min}=0$ is the minimum level of the \gls{BES}.
The \gls{SoC} is the ratio of remaining to nominal maximum stored energy:
\begin{align} \label{subequation:BES-8}
	SoC_{t} = \frac{E_{t}}{E^{\max}}.
\end{align}
The \gls{BES}'s \gls{SoC} ranges from $0~\%$ (fully discharged) to $100~\%$ (fully charged). 
This \gls{SoC} interval may vary depending on the \gls{BES} type used and the manufacturer's specifications.
Discharging the \gls{BES} to a \gls{SoC} below $20~\%$ accelerates the degradation~\cite{Marra_2014}.
To ensure optimal lifespan and operation, the manufacturer prescribes maintaining an \gls{SoC} between $20~\%$ and $90~\%$~\cite{SMA_SmartHome_2023}, which Eq.~(\ref{subequation:BES-9}) defines as follows:
\begin{align} \label{subequation:BES-9}
	SoC^{\min} \leq &SoC_{t}     \leq SoC^{\max}
\end{align}
Finally, Eq. (\ref{subequation:BES-10}) introduces a grid constraint associated with line loading~\cite{Thurner_2018}:
\begin{align} \label{subequation:BES-10}
	\mathcal{L}_{i} = \frac{ \max(i_{\text{from}}, i_{\text{to}} ) }{ i_{\text{thermal}}^{\max} } \cdot 100
\end{align}
% i_ka: Maximum of i_from_ka and i_to_ka [kA]
% i_ka = (i_from_ka, i_to_ka)
% i_ka = (Current at from bus [kA],Current at to bus [kA])
% = Maximum of i_from_ka and i_to_ka [kA]
% max_i_ka* = maximal thermal current [kilo Ampere]
% parallel* = number of parallel line systems
% df (float, 1) - derating factor: maximal current of line in relation to nominal current of line (from 0 to 1)
% loading_percent = ((i_ka)/())
% line loading [%] = (()/())
where \(\mathcal{L}_{i}\) denotes the line loading of the distribution line $i$, expressed as a percentage. 
%This distribution line connects the bus $i$ to its adjacent bus. 
%The line current can be calculated from the maximum of the currents $i_{\text{from}}$ flowing from bus $i$ and $i_{\text{to}}$ flowing to the neighboring bus.
To calculate the line loading, the maximum line current of the connected distribution line $i$ is divided by the specified maximum thermal current~$i_{\text{thermal}}^{\max}$.
\subsection{Influence of the Configurable \gls{MPV} Rates}
\label{sec:Influence of the Configurable MPV Rates}
Based on several characteristics --- $\alpha$ (penetration), $\beta$ (concentration), $\gamma_{1}$ (configurable apparent peak power), and  $\gamma_{2}$ (configurable solar cell power) --- we evaluate the impact of \gls{MPV} on the grid.
This analysis follows the PV assessments described in~\cite{MendezQuezada_2006}. 
These configurable parameters can be used to perform a sensitivity analysis of the influence of different configurations of \gls{MPV} systems on the performance metrics of the \gls{LV} distribution grid.
\subsubsection{$\alpha-Characteristic$: Penetration of \gls{MPV} Systems}
For assessing the influence of \gls{MPV} systems on the \gls{LV} grid, we introduce \gls{MPV} penetration rate \(\alpha\), representing a dimensionless ratio between the \gls{MPV} generation and the consumption of load, defined as:
\begin{align}
\alpha = \frac{E_{\text{MPV Capacity}}(\gamma_{1}, \gamma_{2}, T_{\text{sim}})}{E_{\text{Feeder Capacity}}(T_{\text{sim}})}
\end{align}
where \(E_{\text{MPV Capacity}}(\gamma_{1}, \gamma_{2}, T_{\text{sim}})\) represents the aggregate energy output of \gls{MPV} systems in the distribution grid over the simulation period $T_{\text{sim}}$, \(\gamma_{1}\) describes the configurable apparent peak power of the MPV microinverter, \(\gamma_{2}\) represents the configurable solar cell power of the MPV in the configuration file and \(E_{\text{Feeder Capacity}} (T_{\text{sim}})\) symbolizes the grid's total energy potential over the simulation period $T_{\text{sim}}$.
\subsubsection{$\beta-Characteristic$: Concentration of \gls{MPV} Systems}
Considering the spatial distribution of \gls{MPV} systems in the electrical grid, we define $\beta$ as the concentration rate that is expressed by:
%\vspace*{-1.5mm}
\begin{align}
	\beta = \frac{N_{\text{MPV Buses}}}{N_{\text{Load Buses}}} 
\end{align}
where \( N_{\text{MPV Buses}} \) and \( N_{\text{Load Buses}} \) denote \gls{MPV}-attached buses and total load buses, respectively.
This $\beta$-metric compares the number of MPV systems with the number of loads in the grid.
At each load unit, we install one \gls{MPV}s with concentrations ranging from $0$~$\%$ (indicating no \gls{MPV} installation) to $100$~$\%$ (indicating one \gls{MPV} unit at each load unit, resulting in a full \gls{MPV} concentration level).

	% Section VI: Control Strategies for DER Systems
	\section{Control Strategies for DER Systems}
	\label{sec:InverterControlStrategiesinVDE-AR-N4105}
	% Author(s): Gökhan Demirel
% Author Affiliation:
% Institute for Automation and Applied Informatics (IAI),
% Karlsruhe Institute of Technology (KIT),
% Eggenstein-Leopoldshafen, Germany
%!TEX root = ./Impact_and_Integration_of_Mini_Photovoltaic_Systems_on_Electric_Power_Distribution_Grids.tex
% 006Section01ReactivePowerandVoltageControlforDER.tex
%However, with the increased adoption of \gls{DER}s, the \gls{Q(P, V)} mode is gaining more traction~\cite{Demirok_2011} as an alternative control mode.
DSOs have traditionally managed voltage regulation through grid reinforcement, including measures by adding lines or modernization of transformers~\cite{Carvalho_2008}.
The grid code described in~\cite{VDE_2018} forms the basis of this Section.
\gls{DER} systems mainly use three local reactive power control modes: \gls{Q(V)}, \gls{Q(P)}, and the fixed \gls{cosvarphi} mode, as described in Section \ref{sec:Reactive Power and Voltage Control}.
These control modes, defined by constant values or piecewise first-order curves, allow \gls{DSO}s to change these settings remotely, provided they adhere to local \gls{DER} regulations.
Although \gls{MPV}s are technically usable for voltage regulation, they are not used in this role since \gls{VDE} regulations do not prescribe reactive power control under $1~kV$. 
It is therefore not considered.
To optimize system efficiency, control strategies for \gls{PV}, \gls{BES} and combined \gls{PV}-\gls{BES} systems are considered:
\begin{enumerate}[leftmargin=*]
\item Decentralized grid control: 
Each \gls{DER} unit autonomously makes decisions at the local \gls{PCC}, independent of grid constraints.
\item Distributed grid control: All \gls{DER} units along the feeder control zones coordinate the decision-making process jointly.
\end{enumerate}
%In distributed strategies, \gls{BES} and \gls{PV} systems collaborate within feeder zones for grid support and self-consumption optimization.
%These rule-based inverter control strategies ensure voltage regulation and reactive power management and increase the hosting capacity of \gls{DER} in \gls{LV} grids.
\subsection{Reactive Power and Voltage Control for \gls{DER}}
\label{sec:Reactive Power and Voltage Control}
\subsubsection{\gls{Q(V)}$-Control$: Reactive Power-Voltage Characteristic}
The \gls{Q(V)} -- Control mode primarily addresses the interplay between reactive power and voltage within \gls{DER} units.
The primary design intent of this mode is to oversee and regulate the reactive power interchange between the \gls{DER} unit and the distribution grid.
In this configuration, real-time voltage monitoring occurs at the \gls{PCC} for \gls{DER} units.
To uphold accuracy, the deviation in measurement should not exceed 1~$\%$ of the $p.u.$ value.
A foundational dead band around the fixed reference voltage $V_{\text{ref}}$, together with linear Droop control, establishes the basis for the \gls{Q(V)} characteristic:
\begin{align} \label{Q(V)-equation:DVCS-1} 
\text{\gls{Q(V)}} = 
\begin{cases}
	Q_{\max}, 							        & V \leq V_{1} \\
	Q_{\max} (1-\frac{(V-V_{1})}{(V_{2}-V_{1})}),  & V_{1} < V < V_{2} \\
     0, 									    & V_{2} \leq V \leq V_{3} \\
	-Q_{\max} \frac{(V-V_{3})}{(V_{4}-V_{3})},  & V_{3} < V < V_{4} \\
    -Q_{\max}, 								    & V \geq V_{4}
\end{cases}
\end{align}
Parameters $V_{1}$, $V_{2}$, $V_{3}$, and $V_{4}$ in this piecewise function are adjustable based on \gls{DSO} conditions and set the voltage deadband limits.
\subsubsection{\gls{Q(P)}$-Control$: Active Power-Power Factor Characteristic}
The \gls{Q(P)}$-Control$ mode primarily targets the direct adjustment of the \gls{DER} unit's reactive power in response to its momentary active power generation.
Commonly known as the \gls{PF}, this relationship is symbolized by \gls{cosvarphi}.
It represents the cosine value of the phase difference existing between active and apparent power.
When the measured active power actual value $P_{\text{AV}}$ exceeds the $P_{1}$ threshold, a linear transition in the \gls{cosvarphi}$(P_{\text{AV}})$ value ensues, shifting from a predefined configurable \gls{PF} value \(C_{1}\) to \(C_{2}\).
The characteristic of the curve is illustrated by the following:
\begin{align} \label{Q(P)-equation:DVCS-2}
	\text{\gls{cosvarphi}}(P) = 
	\begin{cases}
		C_{1}, 							    					& P < P_{1} \\
		C_{1} + (P-P_{1})\frac{(C_{2} -C_{1})}{(P_{2}-P_{1})},  & P_{1} \leq P \leq P_{2} \\
		C_{2}, 													& P > P_{2}
	\end{cases}
\end{align}
In mathematical terms, the reactive power can be represented as follows:
\begin{align} \label{Q(P)-equation:DVCS-3}
\text{\gls{Q(P)}} =  P \cdot \tan(\arccos \varphi(P)) 
\end{align}
This formula captures the dynamic interplay of reactive power \gls{Q(P)} and the real-time active power value $P_{\text{AV}}$ using the \gls{PF} characteristic \gls{cosvarphi}$(P)$, as outlined in Eq. (\ref{Q(P)-equation:DVCS-2}).
\subsubsection{$Fixed$ \gls{cosvarphi}$-Control$: Constant Power Factor Characteristic}
In contrast to methods with feedback control mechanisms, a constant ratio of active to reactive power is preferred in the mode \(\textit{Fixed}\) \gls{cosvarphi}\(-\textit{Control} \) without a feedback loop, which is referred to as \gls{cosvarphi}~=~\(\text{const}\).
In this mode, the reactive power from the \gls{DER}, as depicted in Eq.~(\ref{Q(P)-equation:DVCS-3}), correlates directly with its active power, with the interconnection starting at a threshold value of the active power $P_{1}$.
The grid operator determines the fixed setpoint for the factor \gls{cosvarphi} according to the grid code within the permissible range of the \gls{DER} unit.
	% Author(s): Gökhan Demirel
% Author Affiliation:
% Institute for Automation and Applied Informatics (IAI),
% Karlsruhe Institute of Technology (KIT),
% Eggenstein-Leopoldshafen, Germany
%!TEX root = ./Impact_and_Integration_of_Mini_Photovoltaic_Systems_on_Electric_Power_Distribution_Grids.tex
% 0062SectionDecentralizedControlStrategiesforDER.tex
\subsection{Control Strategies for \gls{DER}s}
\label{sec:Decentralized Control Strategies for DER}
We evaluate control strategies for \gls{PV}, \gls{BES}, and combined \gls{PV}-\gls{BES} at the \gls{PCC}. 
Table \ref{tables:strategies} provides an overview of these \gls{DER} control strategies in the \gls{LV} Grid.
In this work, we do not apply these control strategies to \gls{MPV}s.
Active voltage control of \gls{PV} inverters adjusts according to the \gls{VDE} grid code based on \gls{PCC} voltage measurement or actual \gls{PV} power output.
Applying the Dijkstra pathfinding algorithm for the distributed control strategy, we discern the primary branch of each grid, subsequently partitioning control zones between feeders and transformer substations. 
This approach is in line with the requirements of the German Energy Industry Act~\cite{EnWG2005}, which mandates non-discriminatory grid access for all operators and promotes equitable competition within the energy domain.
Notably, some feeder zones may be populated with diverse \gls{DER}s, while others might not, reflecting a realistic grid structure. 
Cost savings in this study derive from minimizing the total active losses, assuming a constant electricity price.
The minimization includes factors such as reactive power generation and grid losses. 
Eq. (\ref{equation:7}) outlines these cost-saving factors that influence overall profitability.
While each \gls{DER} operates in decentral control under individual ownership, the distributed control approach allows multiple owners to collaborate within a feeder zone.
In the distributed control system, the feeder control zones, which extend from the transformer station to the end of the feeder, are equipped with communication devices. 
These devices enable DERs to exchange information and make dynamic decisions on the feed-in or feed-out of power to the grid.
The reactive power control mode described in Section \ref{sec:Reactive Power and Voltage Control} must be used for each \gls{DER} strategy when feeding into the grid. 
\begin{itemize}[leftmargin=*]
	\item \textbf{\gls{PV} Strategy 1 -- \gls{SC}}:	
	The \gls{PV} system primarily powers local consumption in sunny conditions.
	Any residual power, represented as $P_{\text{t}}^{\text{RES}}$, feeds the grid by the difference between \(P_{\text{t}}^{\text{PV}}\) and \(P_{\text{t}}^{\text{Load}}\). 
	If there is a surplus, the power feeds into the grid according to a specified control mode.
	% ---------------------------------BES Strategy 1
	\item \textbf{\gls{BES} Strategy 1 -- \gls{DNC}}:
	The \gls{BES} system operates based on preset charging and discharging cycles dependent on the time of day.
	By default, the system charges from 6:00 a.m. to 6:00 p.m., provided the \gls{SoC} has yet to reach its maximum limit \(SoC^{\max}\).
	Conversely, it discharges from 6:00 p.m. to 6:00 a.m. the next day, as long as the \gls{SoC} remains above its minimum threshold, \(SoC^{\min}\).
	The adaptability of these time configurations enhances its operational flexibility. 
	In the context of the algorithm, $p_{t}^{\text{cha}-\max}$ and $p_{t}^{\text{dis}-\max}$ denote the charging and discharging power of the \gls{BES} system, respectively. \\
	% ---------------------------------PV-BES Strategy 1
	\item \textbf{\gls{PV}-\gls{BES} Strategy 1 -- Decentralized \gls{SC} }:
	Storing excess \gls{PV} energy generated during sunny hours enhances self-consumption.
	This approach gains importance when the feed-in tariff for PV electricity is lower than the purchase price from the grid.
	The primary role of generated \gls{PV} energy is to cover the local load; any surplus energy is stored in the \gls{BES}.
	The residual power, denoted as \(P_{t}^{\text{RES}}\), is the difference between the \gls{PV} generation, \(P_{t}^{\text{PV}}\), and the load demand at the \gls{PCC}, \(P_{t}^{\text{Load}}\). 
	When \gls{PV} generation is available and the \(SoC^{\max}\) limit is reached, excess power is supplied to the grid. Simultaneously, the voltage regulation modes are supported.
	The \gls{BES} primarily ensures self-sufficiency at the \gls{PCC} and regulates only the active power for self-consumption. \\
	% ---------------------------------PV-BES Strategy 2
	\item \textbf{\gls{PV}-\gls{BES} Strategy 2 -- Distributed \gls{SC} }:
	In the distributed \gls{PV}-\gls{BES} strategies, the \gls{LV} grid is segmented into control zones. 
	Each operator manages a feeder control zone from the transformer substation to the end of the feeder.
	For every time step $t$, we calculate the total power consumption and \gls{PV} generation within the zone to determine the residual power \(P_{t, \text{zone}}^{\text{RES}} \). 
	If there is a power surplus and the \gls{SoC}s of the \gls{BES}s are below their maximum levels, the \gls{BES}s charge at a rate dependent on the residual power, and the maximum charging rate \(p^{\text{cha}-\max}\). 
	When the \gls{SoC}s reach their maximum levels, the \gls{PV}s supply surplus power to the grid. 
	Conversely, with a power deficit and the \gls{BES}s above their minimum levels, the \gls{BES}s discharge based on the residual power and the maximum discharging rate \(p^{\text{dis}-\max}\), while sourcing any shortfall from the grid. \\
	% ---------------------------------PV-BES Strategy 3
	\item \textbf{\gls{PV}-\gls{BES} Strategy 3 -- Distributed \gls{SC}-\gls{DNC}}:
	The charging and discharging processes derive from the \gls{DNC} and \gls{SC} strategies by integrating them into a \gls{SC}-\gls{DNC} strategy.
	This time-based operation ensures time-aligned charging/discharging cycles and maximized self-consumption by leveraging \gls{PV} generation.
\end{itemize}
	% Author(s): Gökhan Demirel
% Author Affiliation:
% Institute for Automation and Applied Informatics (IAI),
% Karlsruhe Institute of Technology (KIT),
% Eggenstein-Leopoldshafen, Germany
%!TEX root = ./Impact_and_Integration_of_Mini_Photovoltaic_Systems_on_Electric_Power_Distribution_Grids.tex
% 006Section02ControlStrategiesforDER-tab01.tex
\begin{table}%[h]
	\centering
%	\vspace*{-3mm}
	\caption{ANALYZED DER CONTROL STRATEGIES IN LV GRID}
	\label{tables:strategies}
	\resizebox{\columnwidth}{!}{%
		\begin{tabular}{cl}
			\hline
			\textbf{Strategies} & \textbf{Description} \\ \hline
			\textbf{PV Strategy 1} & 
			\begin{tabular}[l]{@{}l@{}}
				- Powers local consumption \(P_{\text{t}}^{\text{Load}}\) \\
				- Feeds residual power \(P_{\text{t}}^{\text{RES}}\) into the grid \\
			\end{tabular} \\ \hline
			\textbf{BES Strategy 1} & 
			\begin{tabular}[l]{@{}l@{}}
				- Charges BES to \(SoC^{\max}\) from 6:00 a.m. to 6:00 p.m.\\
				- Discharges BES to \(SoC^{\min}\) from 6:00 p.m. to 6:00 a.m.\\
			\end{tabular} \\ \hline
			\textbf{PV-BES Strategy 1} & 
			\begin{tabular}[c]{@{}l@{}}
				- Charges BES if residual local power (\(P_{t}^{\text{RES}} > 0\)) \\
				- Discharges BES if residual local power (\(P_{t}^{\text{RES}} < 0\)) \\
			\end{tabular}  \\ \hline
			\textbf{PV-BES Strategy 2} & 
			\begin{tabular}[c]{@{}l@{}}
				- Each operator manages a control zone within LV grid \\
				- Calculates joint zone residual power \(P_{t, \text{zone}}^{\text{RES}}\) \\
				- Charges BES with residual power if (\(P_{t, \text{zone}}^{\text{RES}} > 0\)) \\
				- Discharges BES with residual power if (\(P_{t, \text{zone}}^{\text{RES}} < 0\)) \\
				- Grid supply if (\(P_{t, \text{zone}}^{\text{RES}} > 0\)) and \(SoC_{t,zone} = SoC^{\max}\) \\
				- BES idle mode if no charging or discharging is required
			\end{tabular}  \\ \hline
			\textbf{PV-BES Strategy 3} & 
			\begin{tabular}[c]{@{}l@{}}
				- Charges BES during DNC periods if (\(P_{t, \text{zone}}^{\text{RES}} > 0\)) \\
				- Discharges BES during DNC periods if (\(P_{t, \text{zone}} ^{\text{RES}} < 0\)) \\
			\end{tabular} \\ \hline
			% Fügen Sie weitere Strategien nach Bedarf hinzu
		\end{tabular}
	}
\end{table}
	
	% Section VII: Case Studies
	\section{Case Studies}
	\label{sec:CaseStudies}
	% Author(s): Gökhan Demirel
% Author Affiliation:
% Institute for Automation and Applied Informatics (IAI),
% Karlsruhe Institute of Technology (KIT),
% Eggenstein-Leopoldshafen, Germany
%!TEX root = ./Impact_and_Integration_of_Mini_Photovoltaic_Systems_on_Electric_Power_Distribution_Grids.tex 
% 007Section01ExperimentalSettings.tex
\subsection{Experimental Settings}
\label{sec:ExperimentalSettings}
This section describes the experimental settings for three case studies in a 15-bus electrical grid.
Case study 1 performs a sensitivity analysis extending over 9600 time steps, corresponding to a duration of 100 days in summer.
This study focuses on the influence of the integration of MPV systems into the electrical grid, focuses on the parameters $\alpha$, $\beta$, $\gamma_{1}$, and $\gamma_{2}$.
Case study 2 compares different methods of reactive power control over 24 hours, divided into 96 time steps.
The distributed control strategy (PV-BES strategy 2) remains fixed while the grid's response to different reactive power control modes is analyzed.
The third case study compares decentralized and distributed grid control strategies over 24 hours with 96-time steps. The \gls{Q(V)} is fixed in this case. 
We focus on analyzing the charging and discharging processes of \gls{BES}, the \gls{SoC} of the \gls{BES}, and the grid limitation on the distribution lines.
Each simulation starts with random initial values set in the configuration file, with a resolution of $\Delta t = 15~\text{min}$ per step.
%The goal of the case studies is to evaluate the advantages and disadvantages of different inverter-based \gls{DER} control methods by the \gls{VDE} grid code, considering different control strategies for \gls{PV}, \gls{BES} and the combination of \gls{PV}-\gls{BES}.

	% Author(s): Gökhan Demirel
% Author Affiliation:
% Institute for Automation and Applied Informatics (IAI),
% Karlsruhe Institute of Technology (KIT),
% Eggenstein-Leopoldshafen, Germany
%!TEX root = ./Impact_and_Integration_of_Mini_Photovoltaic_Systems_on_Electric_Power_Distribution_Grids.tex 
% 007Section01bDataandGridTopologyDescriptions.tex
\subsection{Data and Grid Topology Descriptions}
\subsubsection{Benchmark Electrical Grid Topology}
\label{sec:Power Distribution Benchmark Grid Topology}
\begin{figure}[H]
	\centering
	% Skalierung der Box, in die das Bild eingefügt wird
	\resizebox{9.0cm}{5.0cm}{
		{
			\fontsize{32pt}{16pt}\selectfont% or whatever fontsize you like
			\def\svgwidth{13.5in} 
			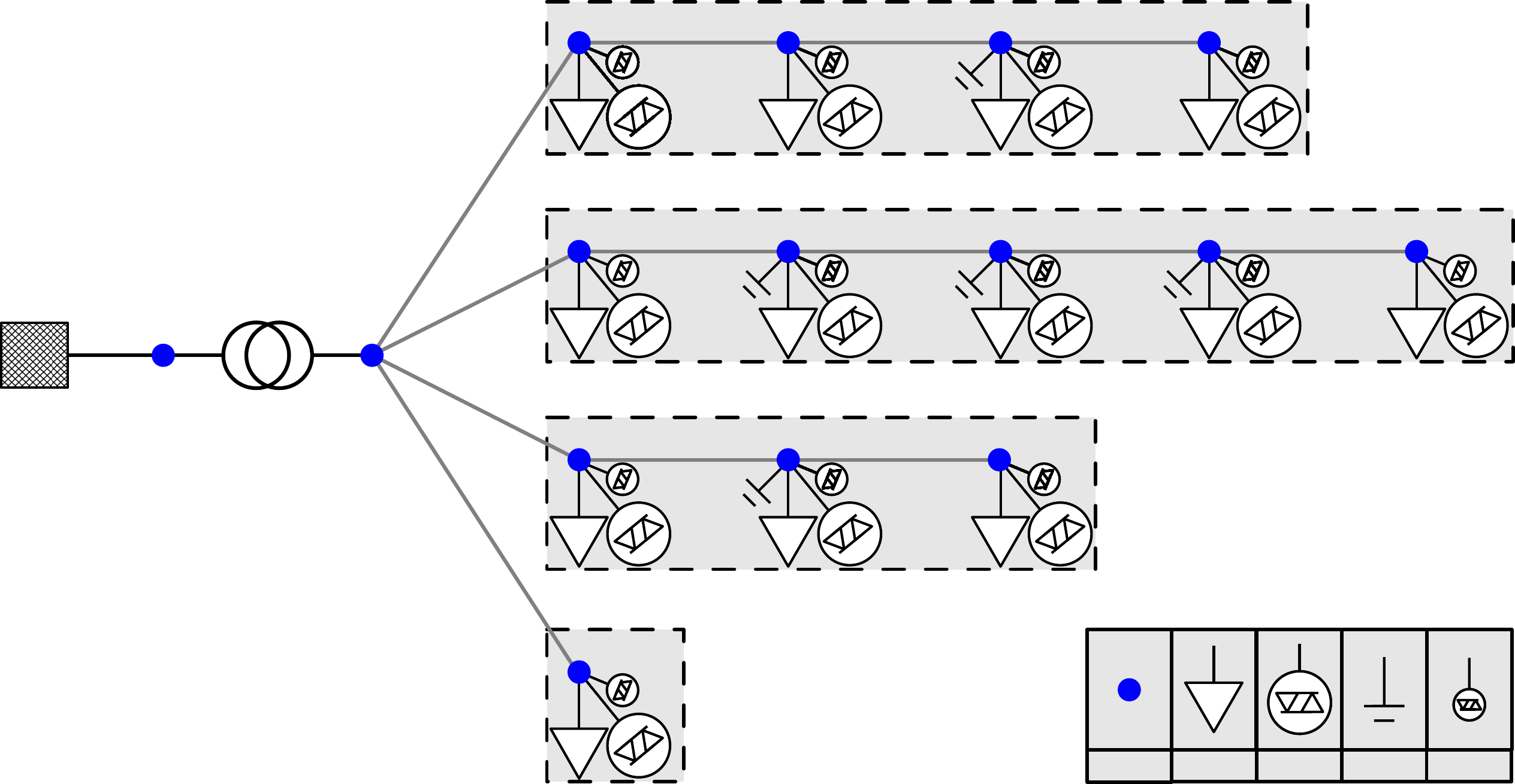
		}
	}
	\caption{\gls{15-bus} \gls{LV} distribution grid segmented into four feeder control zones extending from the feeder end-terminal to the substation. Blue circles mark each bus. Buses 0-4 indicate the main connections between the substation and the external grid.}
	\label{fig:03-13-bus rural control regions}
\end{figure}

To evaluate the control mechanisms in inverter-based DER, we use the benchmark tool \gls{SimBench}\cite{Meinecke_2020}, which provides load and PV profiles as well as electrical grid topologies for the simulation.
These topologies contain all technical grid limitations and installed grid components.
Based on the benchmark, we offer three different configurations of the energy transition:
\begin{itemize}
	\item \gls{Configuration 0 (Base)}: This configuration provides a baseline for the grids and focuses on maintaining voltage stability and adhering to operating standards.
	\item \gls{Configuration 1 (2024)}: This configuration anticipates a significant surge in \gls{DER} deployment.
	\item \gls{Configuration 2 (2034)}: This configuration envisions continued growth in renewable energy adoption driven by increased integration of heat pumps and electric vehicles.
\end{itemize}
We partition the feeder control regions described in Section~\ref{sec:Decentralized Control Strategies for DER} and integrate them into the grid topology information.  
This process, including the partitioning of the control zones in the \gls{LV} grids, is illustrated in Fig.~\ref{fig:03-13-bus rural control regions}, which shows the specific case "1-LV-rural1-2-sw" of the \gls{15-bus} electrical distribution grid topology with the added grid components. 
With a configurable \gls{MPV} concentration, as described in Section \ref{sec:Influence of the Configurable MPV Rates}, additional \gls{MPV}s units can be added to the electrical grid. 
In the general approach for all grid topologies described, we extend each \gls{BES} bus with solar power plants, leading to a \gls{PV}-\gls{BES} system. Specifically, in the case studies of the 15-bus electrical grid, this extension involves adding five solar power plants at buses 6, 9, 10, 12, and 14, as depicted in Fig.~\ref{fig:03-13-bus rural control regions}.
The numbers in black represent the bus indexes.
The grid consists of 15 buses, 13 lines, 1 transformer, 5 \gls{BES}, and 13 \gls{PV} systems, with bus 0 acting as a slack bus.
The number of \gls{MPV}s varies depending on the configuration parameters. The initial state \gls{SoC} of all \gls{BES} on buses 6, 9, 10, 12, 14 is 0$\%$. 
%The \gls{BES} units in control zone 1 (buses 8, 11, 10, and 3), control zone 2 (buses 7, 12, 14, 6, and 5), and control zone 3 (buses 2, 9, and 13) are controlled by the methods listed in Table~\ref{tables:strategies}.
\subsubsection{Load and \gls{PV} Profiles}
In our simulation, load, \gls{PV} and \gls{MPV} power data are added with truncated Gaussian noise, representing noisy measurements.
The standard deviations of this noise are set to $\sigma_{\text{noise}} =  [\sigma_{\text{noise}}^{\text{Load}},\sigma_{\text{noise}}^{\text{\gls{PV}}}, \sigma_{\text{noise}}^{\text{\gls{MPV}}}] = [10^{2}, 10, 10]$ in~\gls{pu} and correspond to the respective data of the grid elements.
The \gls{PV} profile has a \gls{cosvarphi} magnitude of 1.0. 
The \gls{PV} penetration rates are 2.81 for case study 1 and 4.04 for case studies 2 and 3.
%corresponding to the ratio between \gls{PV} generation and load consumption. 
\subsubsection{\gls{MPV} Profiles}
Two regions in Germany, 
\ifblind (blinded for review) \else the urban district of Pforzheim and the rural district of Karlsruhe in the federal state of Baden-Württemberg \fi have different zonal solar radiation characteristics.
Smart plugs record the output data of \gls{MPV} units in real-time as \gls{MPV} profiles.
A total of six \gls{MPV}s are paired at three different locations, with two \gls{MPV}s at each location ensuring spatial-geometric correlation.
For the case studies, the \gls{MPV} units in the same control zone have the same \gls{PV} profiles from the benchmark as they are geometrically correlated.
Each load is randomly assigned an \gls{MPV} unit in the control zones within the electrical grid, and the \gls{MPV} profiles are also randomly assigned.
%This \gls{MPV} profile data are modified based on the configurable parameters such as \gls{MPV} penetration, \gls{MPV} concentration levels, \gls{MPV} peak power, and \gls{MPV} solar cell power in Section~\ref{sec:Influence of the Configurable MPV Rates}.
This paper uses an MPV penetration rate $\alpha$ of 0.53 for case studies 1 and 0.75 for case studies 2 and 3.

	% Author(s): Gökhan Demirel
% Author Affiliation:
% Institute for Automation and Applied Informatics (IAI),
% Karlsruhe Institute of Technology (KIT),
% Eggenstein-Leopoldshafen, Germany
%!TEX root = ./Impact_and_Integration_of_Mini_Photovoltaic_Systems_on_Electric_Power_Distribution_Grids.tex
% fig03.tex
%	Legends indicate grid components including Loads, \gls{PV}s, \gls{MPV}s, and \gls{BES}s.
%\resizebox{8.5cm}{4.5cm}{
%	{
%		\fontsize{10pt}{9pt}\selectfont% or whatever fontsize you like
%		\def\svgwidth{7.333in}
%		\input{figures/pdf/10figControlZones_v5_small.pdf_tex}
%	}
%} % Include figure (example: fig03)
	% Author(s): Gökhan Demirel
% Author Affiliation:
% Institute for Automation and Applied Informatics (IAI),
% Karlsruhe Institute of Technology (KIT),
% Eggenstein-Leopoldshafen, Germany
%!TEX root = ./Impact_and_Integration_of_Mini_Photovoltaic_Systems_on_Electric_Power_Distribution_Grids.tex
% fig04.tex
\begin{figure*}%[H]
	\centering
%	\resizebox{1.00\textwidth}{!}{\input{plots/pgf/combined_4x4_plot_21_11_3.pgf}}
%	\resizebox{1.00\textwidth}{!}{\input{plots/pgf/combined_1st_4x4_plot_21_11.pgf}}
	\resizebox{1.00\textwidth}{!}{\input{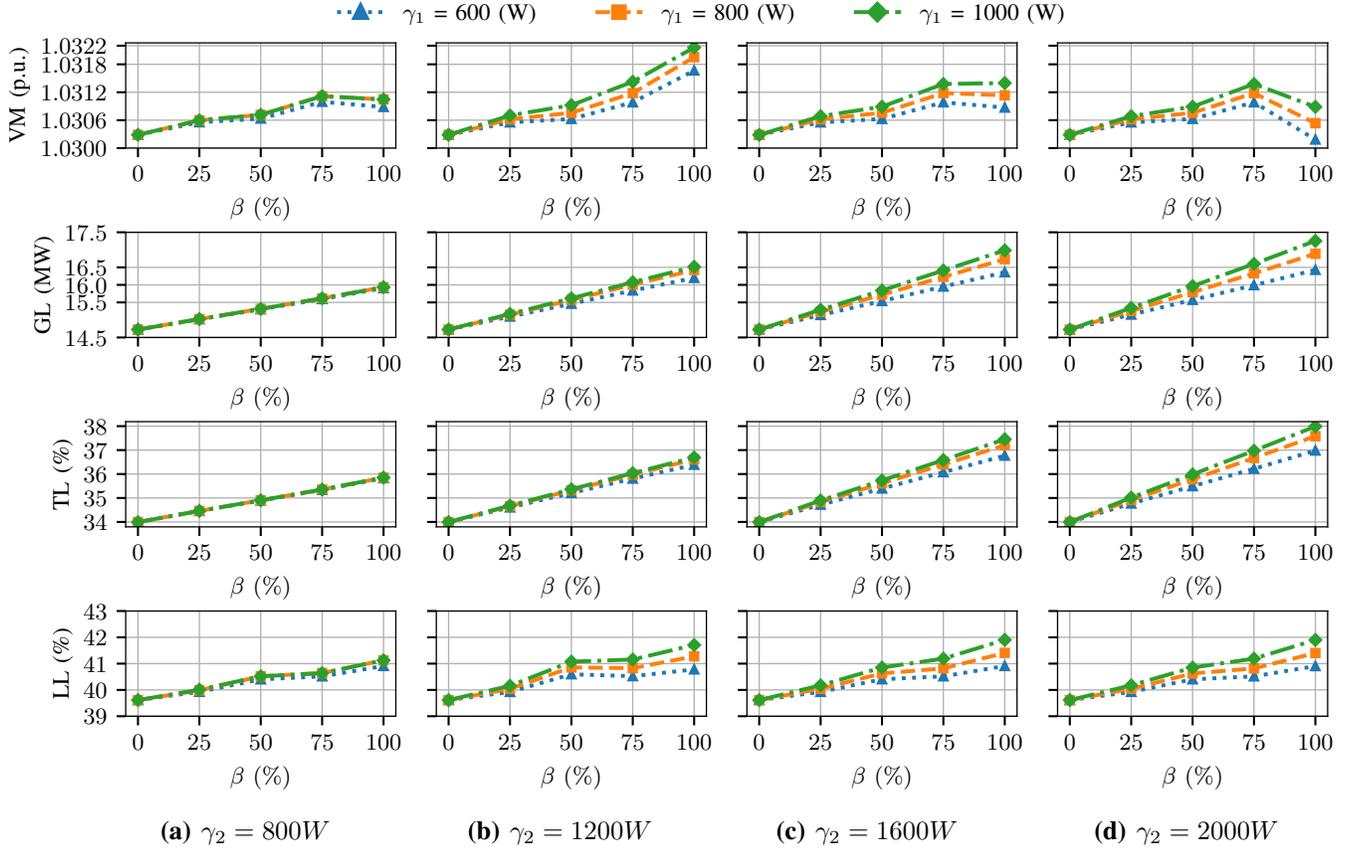}}	
	% .pgf
	%	\input{plots/pgf/combined_4x4_plot_21_14.85.pgf}
	% Manuelle Platzierung der Subfigure-Captions 
	\begin{minipage}[t]{0.30\textwidth}
		\centering
		\textbf{\qquad \qquad (a) $\gamma_{2} = 800 W$}
	\end{minipage}%
	\hfill
	\begin{minipage}[t]{0.20\textwidth}
		\centering
		\textbf{\qquad (b) $\gamma_{2} = 1200 W$}
	\end{minipage}%
	\hfill
	\begin{minipage}[t]{0.25\textwidth}
		\centering
		\textbf{\qquad (c) $\gamma_{2} = 1600 W$}
	\end{minipage}%
	\hfill
	\begin{minipage}[t]{0.25\textwidth}
		\centering
		\textbf{ (d) $\gamma_{2} = 2000 W$}
	\end{minipage}
	\caption{Illustration of the influence of integrating \gls{MPV}s into \gls{LV} grids using four solar cell capacities: subfigure (a) 800~W, (b) 1200~W, (c) 1600~W, and (d) 2000~W. 
	Each subfigure shows the interaction between the \gls{MPV} concentration rate and performance metrics for three different levels of \gls{MPV} inverter apparent power: 600~W, 800~W, and 1000~W.
	}
	\label{fig:13_3fig}
\end{figure*} % Include figure (example: fig04)
	% Author(s): Gökhan Demirel
% Author Affiliation:
% Institute for Automation and Applied Informatics (IAI),
% Karlsruhe Institute of Technology (KIT),
% Eggenstein-Leopoldshafen, Germany
%!TEX root = ./Impact_and_Integration_of_Mini_Photovoltaic_Systems_on_Electric_Power_Distribution_Grids.tex 
% 007Section01cPerformanceMetrics.tex
%[66] R. Brown, “Impact of smart grid on distribution system design,” in Proc. IEEE Power and Energy Society General Meeting, 2008
\vspace*{-20pt}
\subsection{Performance Metrics}
The integration of \gls{MPV} as a renewable energy source into the generation portfolio can cause additional stress on the electrical distribution grids.
Most renewable energy sources are connected to the grid via inverters, which must be technically capable of supplying reactive power.
In practice, \gls{MPV}s can provide reactive power, but this function is not taken into account by the regulatory framework.
Considering the inherent limitations of distribution lines, such as non-negligible resistances and limited transmission capacities, overload problems, including the number of violations of overload limits, the average overload of lines and transformers, voltage magnitudes, aggregate reactive power injection from DER inverters and power losses, can be used to evaluate grid control methods~\cite{Brown_2008, Dorfler_2019, Mueller_2023}.
We introduce four different performance metrics to evaluate the case studies of our proposed inverter control strategies:
\begin{itemize}[leftmargin=*]
	\item \textbf{\gls{VM}:} 
	This metric calculates the average \gls{VM} across all buses for each simulation step, expressed in~\gls{pu}.
	\item \textbf{\gls{GL}:} 
	This metric calculates the total losses of lines and transformers across all buses for each simulation step. The \gls{GL} are aggregated over time in MW.
	\item \textbf{\gls{TL}:} 
	This metric calculates the average \gls{TL} for each time step during a simulation and represents the load utilization in relation to the rated power as a percentage.
	\item \textbf{\gls{LL}:}
	This metric calculates the average \gls{LL} per time step during a simulation. Eq. (\ref{subequation:BES-10}) provides it as a percentage.
\end{itemize}
\vspace*{-10pt}
%By evaluating these metrics, one could anticipate potential reductions in grid losses and alleviation of stress on system components and ensure cost savings.
%This evaluation methodology aims to identify the optimal control mode tailored for each configuration and topology, prioritizing high \gls{CVR} and minimized \gls{PL}.
%Consequently, the overarching impact lies in economic incentives like grid loss mitigation, promoting the maintenance of minimal reactive power, and ensuring that grid components operate outside their overload range --- representing one of the most significant effects.
	% Author(s): Gökhan Demirel
% Author Affiliation:
% Institute for Automation and Applied Informatics (IAI),
% Karlsruhe Institute of Technology (KIT),
% Eggenstein-Leopoldshafen, Germany
%!TEX root = ./Impact_and_Integration_of_Mini_Photovoltaic_Systems_on_Electric_Power_Distribution_Grids.tex
% 007Section01cGridControlEvaluationEnvironment.tex
\subsection{Grid Control Evaluation Environment}
This subsection introduces our environment, building on the established electrical grid analysis tool \gls{PandaPower}~\cite{Thurner_2018} using a Newton-Raphson power flow solver.
Our modular open-source environment provides a versatile tool for researchers engaged in implementing, replicating, and benchmarking versatile task strategies for system service provision, which includes reactive power control and grid management for \gls{DER}s, adhering to the guidelines for benchmark environments~\cite{2020_wolfle}.
It operates as a discrete-time decision process and outputs observations and rewards, including the results of the power flow and performance metrics shown in the case studies.
Our environment integrates five key characteristics for suitable environments as outlined in~\cite{2020_wolfle}: scenario, relevance, scope, realism, and reproducibility.
We focus on modeling electrical grids with stochastic uncertainties for the scenario feature, on applying reactive power control methods according to the grid code for relevance feature, and on evaluating controllers through decentralized and distributed strategies for \gls{DER}s to define the scope and the reproducibility characteristics.
In addition, grid analysis framework ensures the realism feature.
Case studies demonstrate various reactive power control modes and grid control algorithms for \gls{DER}s, reflecting the environment's capability to handle different tasks and algorithm performance evaluation, as required by~\cite{Beiranvand2017} and~\cite{2020_wolfle}.
	\subsection{Results}
	% Author(s): Gökhan Demirel
% Author Affiliation:
% Institute for Automation and Applied Informatics (IAI),
% Karlsruhe Institute of Technology (KIT),
% Eggenstein-Leopoldshafen, Germany
%!TEX root = ./Impact_and_Integration_of_Mini_Photovoltaic_Systems_on_Electric_Power_Distribution_Grids.tex 
% 007Section04_01_CaseStudy1.tex
%\subsubsection{Case Study 1}
\textbf{Case Study 1:}
In the first case study, we analyze the sensitivity considering the parameters $\alpha$, $\beta$, $\gamma_{1}$, and $\gamma_{2}$, as shown in Fig.~\ref{fig:13_3fig}.
In this case, we only use the distributed grid control strategy with the \gls{PV}-BES strategy 3 from Table \ref{tables:strategies} and the reactive power control \gls{Q(V)} from Section \ref{sec:Reactive Power and Voltage Control}. 
This sensitivity analysis addresses how the integration of \gls{MPV} influences the electrical grid depending on the parameterization, particularly regarding the penetration and concentration of \gls{MPV} as well as the configurable peak and solar cell power.
This allows a deeper understanding of these key parameters impact on the influence of \gls{MPV} on the grid.
As indicated in the legend, the blue dashed line with the triangular marker represents the configurable apparent power peak ($\gamma_{1}$) for $600$~W; the orange dashed line with the square marker for $800$~W and the green dotted-dashed lines with the diamond marker for $1000$~W.
The horizontal axis represents the concentration level of the \gls{MPV}, $\beta$, in percentage, while the vertical axis shows the respective key performance metrics according to their corresponding units.
The subfigures represent the solar cell capacity parameter $\gamma_{2}$ for (a) $800$~W, (b) $1200$~W, (c) $1600$~W and (d) $2000$~W.
The results show that the average \gls{VM} for $\gamma_{2}=1200$ shows a piecewise linear increase with increasing $\beta$.
This phenomenon can be attributed to the reactive power control of \gls{DER} systems.
The analysis shows that both the \gls{TL}, which varies from an average of $34$~$\%$ to $38$~$\%$ and the \gls{LL}, which increases from $39.5$~$\%$ to $42$~$\%$, follow a piecewise linear function in response to increasing parameter values of $\beta$ and $\gamma_{2}$.
Due to the pure active power feed-in of the \gls{MPV}, the integration significantly increases the active power injection at the \gls{PCC}. \\
To sum up, the sensitivity analysis shows that \gls{DER}s counteract this with reactive power control to keep the voltages within limits.
As a result, the integration of the \gls{MPV} leads to an increase in the reactive power provision and inverter losses. 
In addition, more reactive power leads to higher inverter losses~\cite{Braun_2009}.
This study neglects deviations in active power loss between inverters due to different \gls{PF} settings. \\

	% Author(s): Gökhan Demirel
% Author Affiliation:
% Institute for Automation and Applied Informatics (IAI),
% Karlsruhe Institute of Technology (KIT),
% Eggenstein-Leopoldshafen, Germany
%!TEX root = ./Impact_and_Integration_of_Mini_Photovoltaic_Systems_on_Electric_Power_Distribution_Grids.tex
% fig05.tex

\begin{figure}
	\vspace*{-19mm}
	\hspace*{2mm}
	\centering
	\resizebox{1.24\columnwidth}{!}{\input{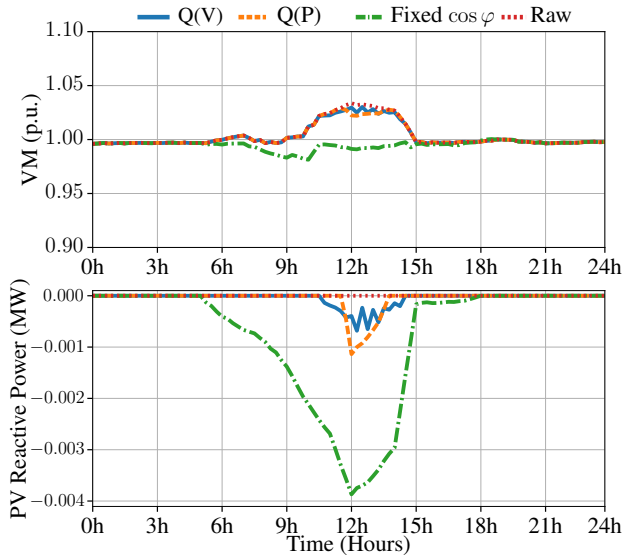}}
	\vspace*{-3mm}
	\caption{Reactive power generation and voltage variation performance in the distribution grid using the different methods on a summer day.}
	\label{fig:Reactive_Power_Control_Modes_fig}
\end{figure}
%\vspace*{-3mm} % Include figure (example: fig05)
	% Author(s): Gökhan Demirel
% Author Affiliation:
% Institute for Automation and Applied Informatics (IAI),
% Karlsruhe Institute of Technology (KIT),
% Eggenstein-Leopoldshafen, Germany
%!TEX root = ./Impact_and_Integration_of_Mini_Photovoltaic_Systems_on_Electric_Power_Distribution_Grids.tex
% fig06.tex
\vspace*{-4mm}
\begin{figure}
%	\vspace*{-1mm}
	\centering
	\resizebox{1.05\linewidth}{!}{\input{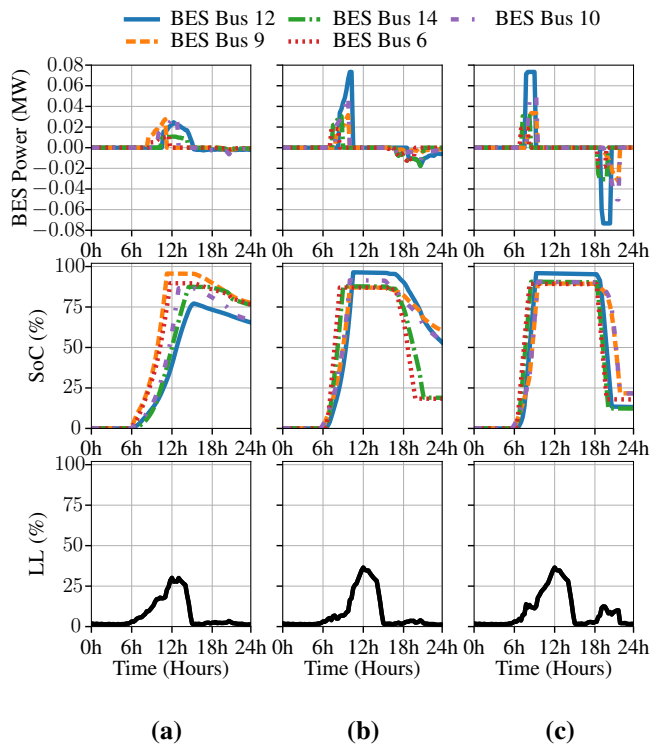}}
	\begin{minipage}[t]{0.33\columnwidth}
		\centering
		\textbf{\qquad\qquad (a)}
	\end{minipage}%
	\hfill
	\begin{minipage}[t]{0.33\columnwidth}
		\centering
		\textbf{\quad\quad(b)}
	\end{minipage}%
	\hfill
	\begin{minipage}[t]{0.33\columnwidth}
		\centering
		\textbf{(c)}
	\end{minipage}%
	\caption{Case Study 3 -- Top subfigures: \gls{BES} charging/discharging under \gls{PV}-\gls{BES} Strategies (a) 1, (b) 2, (c) 3. Middle: Daily \gls{SoC} status. Bottom: Average \gls{LL} status.}
	\label{fig:16_3fig}
\end{figure} % Include figure (example: fig06)
	% Author(s): Gökhan Demirel
% Author Affiliation:
% Institute for Automation and Applied Informatics (IAI),
% Karlsruhe Institute of Technology (KIT),
% Eggenstein-Leopoldshafen, Germany
%!TEX root = ./Impact_and_Integration_of_Mini_Photovoltaic_Systems_on_Electric_Power_Distribution_Grids.tex 
% 007Section04_02_CaseStudy2.tex

%\subsubsection{Case Study 2}
\textbf{Case Study 2:}
In this second case study, we compare different reactive power control modes in our environment while maintaining an unchanged distributed control strategy (\gls{PV}-\gls{BES} strategy 2). Our environment can select and extend different voltage regulation modes, which are then jointly evaluated.
Fig.~\ref{fig:Reactive_Power_Control_Modes_fig} illustrates the reactive power control using different modes: \gls{Q(V)} is shown with a solid blue line, \gls{Q(P)} with a dashed orange line, $\textit{Fixed}$ \gls{cosvarphi} with a green dashed line and no control with a red dotted line.
The first row shows the average \gls{VM} and the second the reactive power feed-in of the \gls{PV} systems. 
Each method controls the voltage within the safety range, despite 100~$\%$~\gls{MPV} concentration levels.
In addition, the \gls{VM} values of \gls{Q(V)} are below those of the $\textit{Fixed}$ \gls{cosvarphi} control but above those of \gls{cosvarphi}.
This phenomenon is possible because the $\textit{Fixed}$ \gls{cosvarphi} control uses a constant \gls{PF}, which cannot explicitly reduce the injected reactive power, and \gls{cosvarphi} injects the reactive power based on a threshold value of the active power.
In contrast, \gls{Q(V)} injects the reactive power only when the threshold voltage is reached, resulting in a \gls{VM} zigzag pattern.
Using \gls{Q(V)} significantly improves \gls{MPV} integration in inverter-based \gls{DER}s, minimizing the curtailment of \gls{DER} active power. 
	% Author(s): Gökhan Demirel
% Author Affiliation:
% Institute for Automation and Applied Informatics (IAI),
% Karlsruhe Institute of Technology (KIT),
% Eggenstein-Leopoldshafen, Germany
%!TEX root = ./Impact_and_Integration_of_Mini_Photovoltaic_Systems_on_Electric_Power_Distribution_Grids.tex 
% 007Section04_03_CaseStudy3.tex
%\subsubsection{Case Study 3}

\textbf{Case Study 3:}
This third case study evaluates decentralized versus distributed grid control strategies, as shown in Table \ref{tables:strategies}.
For all strategies, \gls{Q(V)} was selected as the reactive power control mode.
Fig.~\ref{fig:16_3fig} shows the charging and discharging process of the power in the first row, the \gls{SoC} of the respective \gls{BES} in the second row, and the average \gls{LL} of the distribution lines in the third row.
The subfigures represent (a) \gls{PV}-\gls{BES} strategy 1, (b) strategy 2, and (c) strategy 3.
High solar power generation in the grid leads to a high average load on the lines (\gls{LL}).
In addition, the \gls{LL} increases when the \gls{BES} are actively charging and discharging.
At distributed \gls{DNC} control, this becomes clear between 6:00 p.m. and midnight, when the \gls{BES} units are discharging.
Under the decentralized approach, the \gls{LL} also increases during the charging processes from 6 a.m. to 12 p.m.
The five \gls{BES} units of the distributed \gls{PV}-\gls{BES} strategy 2 charges almost fully between in the same period.
Due to the limitations of \gls{SoC} between 20~$\%$ and 90~$\%$, it is not feasible to charge above 86~$\%$ or discharge below 24~$\%$.
Distributed \gls{DNC} starts charging all \gls{BES} at 6 a.m., based on the charging timestep setting.
In contrast, with the decentralized approach and distributed strategy 2, only one \gls{BES} is fully charged during this period, while the other \gls{BES}s reach a lower \gls{SoC}. 
The \gls{DNC} begins to discharge fully at 6 p.m., causing a high \gls{LL}.
The \gls{BES} on bus 9 is fast charging in decentralized grid control. This fast charging with \gls{MPV} leads on Bus 9 to a moderate increase in \gls{LL}.
This is due to the different load consumption and the lower \gls{BES} capacities.
In decentralized control, which only relies on the local \gls{PV} power, the \gls{SoC} of the four \gls{BES} increases more slowly.
As a result, other \gls{PV} surplus energy actively transfers to the higher-level grid through the slack bus.
However, distributed control strategies enable faster charging as they can exchange access to \gls{PV} power in the feeder control zone. 
%This means that surplus power from other buses is absorbed by the \gls{BES}.
The distributed control system stores energy to compensate for future load peaks, while the decentralized control only covers local demand.
%This case study illustrates the effectiveness of control strategies and emphasizes the versatility and robustness of our environment in simulating real-world grid scenarios.
	
	% Section VIII: Conclusion
	\section{Conclusion}
	\label{sec:Conclusion}
	% Author(s): Gökhan Demirel
% Author Affiliation:
% Institute for Automation and Applied Informatics (IAI),
% Karlsruhe Institute of Technology (KIT),
% Eggenstein-Leopoldshafen, Germany
%!TEX root = ./Impact_and_Integration_of_Mini_Photovoltaic_Systems_on_Electric_Power_Distribution_Grids.tex 
% 008SectionConclusion.tex

The present paper provides a comprehensive analysis of \gls{MPV} systems, commonly known as balcony power plants, and their impact on the stability and control of the \gls{LV} grid. 
Our case studies highlight the essential role of autonomous inverters in providing ancillary services in \gls{DER}-rich grids and underscores the necessity of adaptable \gls{DER} control strategies.
Increasing the use of \gls{DER} in \gls{LV} grids requires new types of reactive power and adaptive control strategies.
Our research advances the understanding of \gls{MPV} systems and their influence on the \gls{LV} grid.
An insightful analysis of the impact of \gls{MPV} systems on technical grid limitations, as well as an evaluation of the response of inverter-based \gls{DER}s to these challenges by compensating higher reactive power, is the focus of this work.
%To sum up, a regulatory framework that protects the grid constraints and potential risks of machine learning under changing environments needs to be developed. 
%The proposed regulatory framework should incorporate adaptive or multi-level control mechanisms that ensure grid stability and minimize active power loss while adhering to present grid codes and safely integrating machine learning algorithms.
 
An open-source Python environment for researching and developing reactive power and grid control strategies has therefore been developed (blinded for review).
The energy community can use this environment to challenge voltage control problems as a fundamental reference and contribute further.
The \gls{MPVBench} dataset, an essential piece of real-time balcony power plant data, is also available on our GitHub repository (blinded for review).
Our future research will focus on integrating other \gls{DER} components, such as heat pumps and electric vehicles, with a focus on AI and \gls{RL} grid control for \gls{DER}s. 
This will enhance our deeper insights into AI-managed renewable energy solutions.

%Our future research will focus on integrating other \gls{DER} components, such as heat pumps and electric vehicles. 
%The primary research direction remains the application of AI and \gls{RL} grid control for \gls{DER}s. 
%This approach will offer deeper insights into the complexities of managing AI-driven solutions in a renewable energy-dominated future.
	
	% Acknowledgment
%	\section*{Acknowledgment}
%	A special thanks to the anonymous reviewers for their insightful comments and pertinent feedback.
	
	% References
	\bibliographystyle{IEEEtran}
	\bibliography{tmp/main/references} % Assuming your BibTeX file is named references.bib

% Generated by IEEEtran.bst, version: 1.14 (2015/08/26)
\begin{thebibliography}{10}
\providecommand{\url}[1]{#1}
\csname url@samestyle\endcsname
\providecommand{\newblock}{\relax}
\providecommand{\bibinfo}[2]{#2}
\providecommand{\BIBentrySTDinterwordspacing}{\spaceskip=0pt\relax}
\providecommand{\BIBentryALTinterwordstretchfactor}{4}
\providecommand{\BIBentryALTinterwordspacing}{\spaceskip=\fontdimen2\font plus
\BIBentryALTinterwordstretchfactor\fontdimen3\font minus \fontdimen4\font\relax}
\providecommand{\BIBforeignlanguage}[2]{{%
\expandafter\ifx\csname l@#1\endcsname\relax
\typeout{** WARNING: IEEEtran.bst: No hyphenation pattern has been}%
\typeout{** loaded for the language `#1'. Using the pattern for}%
\typeout{** the default language instead.}%
\else
\language=\csname l@#1\endcsname
\fi
#2}}
\providecommand{\BIBdecl}{\relax}
\BIBdecl

\bibitem{Tomik_2023_Balkonkraftwerk}
S.~Tomik, \emph{Balkonkraftwerk: Strom selbst erzeugen mit Steckersolarger{\"a}ten und Photovoltaik auf Balkon, Terrasse und im Garten [Balcony Power Plant: Generate Your Own Electricity with Plug-in Solar Devices and Photovoltaics on Balcony, Terrace, and in the Garden]}.\hskip 1em plus 0.5em minus 0.4em\relax Stuttgart, Germany: Ulmer Eugen Verlag, Apr. 2023.

\bibitem{BMWK_2023}
\BIBentryALTinterwordspacing
{Federal Ministry for Economic Affairs and Climate Action (BMWK)}, ``{Photovoltaic Strategy: Fields of Action and Measures for an Accelerated Expansion of Photovoltaics},'' \emph{{Federal Ministry for Economic Affairs and Climate Action}}, vol.~1, no.~1, pp. 1--42, May 2023. [Online]. Available: \url{https://www.bmwk.de/Redaktion/DE/Publikationen/Energie/photovoltaik-stategie-2023}
\BIBentrySTDinterwordspacing

\bibitem{EU2016_631}
\BIBentryALTinterwordspacing
{European Commission}, ``{Commission Regulation (EU) 2016/631 of 14 April 2016 establishing a network code on requirements for grid connection of generators},'' \emph{Official Journal of the European Union}, vol.~1, no. 631, pp. 1--68, April 2016. [Online]. Available: \url{http://data.europa.eu/eli/reg/2016/631/oj}
\BIBentrySTDinterwordspacing

\bibitem{VDE_2023}
\BIBentryALTinterwordspacing
A.~Roth, H.~Kühlmeyer, A.~Nollau, H.~Schäfer, D.~Schädel, and K.~Kreß. (2023, jan) {Plug-and-Play Mini Power Generation Plants}. VDE Association for Electrical, Electronic and Information Technologies. Frankfurt am Main. [Online]. Available: \url{https://www.vde.com/resource/blob/2229846/acbd1078371f6a553a049a1d33b8612c/positionspapier-data.pdf}
\BIBentrySTDinterwordspacing

\bibitem{Poullikkas_2013}
A.~Poullikkas, G.~Kourtis, and I.~Hadjipaschalis, ``A review of net metering mechanism for electricity renewable energy sources,'' \emph{International Journal of Energy and Environment}, vol.~4, pp. 975--1002, 01 2013.

\bibitem{Gautier_2018}
A.~Gautier, J.~Jacqmin, and J.-C. Poudou, ``The prosumers and the grid,'' \emph{Journal of Regulatory Economics}, vol.~53, no.~1, pp. 100--126, jan 2018.

\bibitem{dsire_2023}
\BIBentryALTinterwordspacing
D.~of~State Incentives~for Renewables and E.~(DSIRE). (2023) Net metering policies. dsireusa.org. [Online]. Available: \url{https://www.dsireusa.org/resources/detailed-summary-maps/net-metering-policies-2/}
\BIBentrySTDinterwordspacing

\bibitem{Bergner_2022}
\BIBentryALTinterwordspacing
J.~Bergner, R.~Hoelger, and B.~Praetorius, ``{The Market for Plug-In Solar Devices},'' {Hochschule für Technik und Wirtschaft HTW Berlin}, Tech. Rep.~1, May 2022. [Online]. Available: \url{https://solar.htw-berlin.de/studien/marktstudie-steckersolar-2022}
\BIBentrySTDinterwordspacing

\bibitem{BNetzA_2023}
{Bundesnetzagentur (BNetzA)}. (2024, Jan.) {Archived EEG feed-in tariffs}. Bundesnetzagentur. Bonn. [Quarterly reported new installations of PV systems and current feed-in tariffs of the German Renewable Energy Act] (in German). Bundesnetzagentur. Archived from the \href{https://web.archive.org/web/20240202124405/https://www.bundesnetzagentur.de/DE/Fachthemen/ElektrizitaetundGas/ErneuerbareEnergien/EEG_Foerderung/Archiv_VergSaetze/start.html}{original} on Jan. 2024. Retrieved Feb. 2, 2024.

\bibitem{Geth_2010}
F.~Geth, J.~Tant, E.~Haesen, J.~Driesen, and R.~Belmans, ``Integration of energy storage in distribution grids,'' in \emph{{IEEE PES General Meeting}}.\hskip 1em plus 0.5em minus 0.4em\relax Minneapolis: {IEEE}, July 2010, pp. 1--6.

\bibitem{Thurner_2018}
L.~Thurner, A.~Scheidler, F.~Schafer, J.-H. Menke, J.~Dollichon, F.~Meier, S.~Meinecke, and M.~Braun, ``Pandapower{\textemdash}an open-source python tool for convenient modeling, analysis, and optimization of electric power systems,'' \emph{{IEEE} Transactions on Power Systems}, vol.~33, no.~6, pp. 6510--6521, Nov. 2018.

\bibitem{Meinecke_2020}
S.~Meinecke, D.~Sarajli{\'{c}}, S.~R. Drauz, A.~Klettke, L.-P. Lauven, C.~Rehtanz, A.~Moser, and M.~Braun, ``{SimBench}{\textemdash}a benchmark dataset of electric power systems to compare innovative solutions based on power flow analysis,'' \emph{Energies}, vol.~13, no.~12, p. 3290, June 2020.

\bibitem{Omran_2011}
W.~A. Omran, M.~Kazerani, and M.~M.~A. Salama, ``Investigation of methods for reduction of power fluctuations generated from large grid-connected photovoltaic systems,'' \emph{{IEEE} Transactions on Energy Conversion}, vol.~26, no.~1, pp. 318--327, March 2011.

\bibitem{stetzthomas_2015}
T.~Stetz, J.~von Appen, F.~Niedermeyer, G.~Scheibner, R.~Sikora, and M.~Braun, ``Twilight of the grids: The impact of distributed solar on germany?s energy transition,'' \emph{IEEE Power and Energy Magazine}, vol.~13, no.~2, pp. 50--61, 2015.

\bibitem{Masters_2002}
C.~Masters, ``Voltage rise: the big issue when connecting embedded generation to long 11 {kV} overhead lines,'' \emph{Power Engineering Journal}, vol.~16, no.~1, pp. 5--12, Feb. 2002.

\bibitem{Kerber_2009}
G.~Kerber, R.~Witzmann, and H.~Sappl, ``Voltage limitation by autonomous reactive power control of grid connected photovoltaic inverters,'' in \emph{2009 Compatibility and Power Electronics}.\hskip 1em plus 0.5em minus 0.4em\relax Badajoz: {IEEE}, May 2009, pp. 129--133.

\bibitem{Ari_2011}
G.~Ari and Y.~Baghzouz, ``Impact of high {PV} penetration on voltage regulation in electrical distribution systems,'' in \emph{{International Conference on Clean Electrical Power (ICCEP)}}.\hskip 1em plus 0.5em minus 0.4em\relax Ischia: {IEEE}, June 2011, pp. 744--748.

\bibitem{Tonkoski_2012}
R.~Tonkoski, D.~Turcotte, and T.~H.~M. El-Fouly, ``Impact of high {PV} penetration on voltage profiles in residential neighborhoods,'' \emph{{IEEE} Transactions on Sustainable Energy}, vol.~3, no.~3, pp. 518--527, July 2012.

\bibitem{Stetz_2013}
T.~Stetz, F.~Marten, and M.~Braun, ``Improved low voltage grid-integration of photovoltaic systems in germany,'' \emph{{IEEE} Transactions on Sustainable Energy}, vol.~4, no.~2, pp. 534--542, April 2013.

\bibitem{von_Appen_2014}
J.~von Appen, T.~Stetz, M.~Braun, and A.~Schmiegel, ``Local voltage control strategies for {PV} storage systems in distribution grids,'' \emph{{IEEE} Transactions on Smart Grid}, vol.~5, no.~2, pp. 1002--1009, March 2014.

\bibitem{Stetz_2014}
T.~Stetz, K.~Diwold, M.~Kraiczy, D.~Geibel, S.~Schmidt, and M.~Braun, ``Techno-economic assessment of voltage control strategies in low voltage grids,'' \emph{{IEEE} Transactions on Smart Grid}, vol.~5, no.~4, pp. 2125--2132, July 2014.

\bibitem{Kabir_2014}
M.~N. Kabir, Y.~Mishra, G.~Ledwich, Z.~Y. Dong, and K.~P. Wong, ``Coordinated control of grid-connected photovoltaic reactive power and battery energy storage systems to improve the voltage profile of a residential distribution feeder,'' \emph{{IEEE} Transactions on Industrial Informatics}, vol.~10, no.~2, pp. 967--977, May 2014.

\bibitem{Zeraati_2018}
M.~Zeraati, M.~E.~H. Golshan, and J.~M. Guerrero, ``Distributed control of battery energy storage systems for voltage regulation in distribution networks with high {PV} penetration,'' \emph{{IEEE} Transactions on Smart Grid}, vol.~9, no.~4, pp. 3582--3593, July 2018.

\bibitem{Vargas_2018}
M.~C. Vargas, M.~A. Mendes, and O.~E. Batista, ``Impacts of high {PV} penetration on voltage profile of distribution feeders under brazilian electricity regulation,'' in \emph{{2018 13th IEEE International Conference on Industry Applications (INDUSCON)}}.\hskip 1em plus 0.5em minus 0.4em\relax Sao Paulo: {IEEE}, Nov. 2018, pp. 38--44.

\bibitem{Demirok_2011}
E.~Demirok, P.~C. Gonz{\'{a}}lez, K.~H.~B. Frederiksen, D.~Sera, P.~Rodriguez, and R.~Teodorescu, ``Local reactive power control methods for overvoltage prevention of distributed solar inverters in low-voltage grids,'' \emph{{IEEE} Journal of Photovoltaics}, vol.~1, no.~2, pp. 174--182, Oct. 2011.

\bibitem{Matkar_2017}
G.~Matkar, D.~K. Dheer, A.~S. Vijay, and S.~Doolla, ``A simple mathematical approach to assess the impact of solar {PV} penetration on voltage profile of distribution network,'' in \emph{{National Power Electronics Conference (NPEC)}}.\hskip 1em plus 0.5em minus 0.4em\relax Pune: {IEEE}, Dec. 2017, pp. 209--214.

\bibitem{Singh_2020}
N.~K. Singh, A.~Elrayyah, and M.~Z.~C. Wanik, ``Analysis of voltage rise and optimal {PV} curtailment strategy for its mitigation,'' in \emph{{2020 IEEE PES Innovative Smart Grid Technologies Europe (ISGT-Europe)}}.\hskip 1em plus 0.5em minus 0.4em\relax The Hague: {IEEE}, Oct. 2020, pp. 610--614.

\bibitem{Biel_2018}
D.~Biel and J.~M. Scherpen, ``Active and reactive power regulation in single-phase pv inverters,'' in \emph{2018 European Control Conference (ECC)}.\hskip 1em plus 0.5em minus 0.4em\relax IEEE, Jun. 2018.

\bibitem{Wang_2018}
J.~Wang, F.~Luo, Z.~Ji, Y.~Sun, B.~Ji, W.~Gu, and J.~Zhao, ``An improved hybrid modulation method for the single-phase h6 inverter with reactive power compensation,'' \emph{IEEE Transactions on Power Electronics}, vol.~33, no.~9, pp. 7674--7683, Sep. 2018.

\bibitem{DIN_EN50160_2020}
\BIBentryALTinterwordspacing
{German Institute for Standardisation Registered Association (DIN) and European Committee for Electrotechnical Standardization (CENELEC)}, ``{DIN EN 50160:2022-10, Voltage characteristics of electricity supplied by public distribution networks},'' \emph{DIN Standards}, vol.~2, no.~1, pp. 1--116, Oct. 2022. [Online]. Available: \url{https://dx.doi.org/10.31030/3383427}
\BIBentrySTDinterwordspacing

\bibitem{VDE_2018}
{VDE FNN}, ``{Generators connected to the low-voltage distribution network - Technical requirements for the connection to and parallel operation with low-voltage distribution networks},'' \emph{VDE-AR-N 4105:2018-11 (Revision of VDE-AR-N 4105:2011-08)}, vol.~2, no.~1, pp. 1--96, Nov. 2018.

\bibitem{Subhonmesh_2012}
B.~Subhonmesh, S.~H. Low, and K.~M. Chandy, ``Equivalence of branch flow and bus injection models,'' in \emph{2012 50th Annual Allerton Conference on Communication, Control, and Computing (Allerton)}.\hskip 1em plus 0.5em minus 0.4em\relax Monticello: {IEEE}, Oct. 2012, pp. 1893--1899.

\bibitem{Farivar_2013}
M.~Farivar and S.~H. Low, ``Branch flow model: Relaxations and convexification{\textemdash}part i,'' \emph{{IEEE} Transactions on Power Systems}, vol.~28, no.~3, pp. 2554--2564, Aug. 2013.

\bibitem{Akinyemi_2022}
A.~S. Akinyemi, K.~Musasa, and I.~E. Davidson, ``Analysis of voltage rise phenomena in electrical power network with high concentration of renewable distributed generations,'' \emph{Scientific Reports}, vol.~12, no.~1, pp. 1--22, May 2022.

\bibitem{Baran_1989_sizing}
M.~Baran and F.~Wu, ``Optimal sizing of capacitors placed on a radial distribution system,'' \emph{{IEEE} Transactions on Power Delivery}, vol.~4, no.~1, pp. 735--743, 1989.

\bibitem{Turitsyn_2011}
K.~Turitsyn, P.~Sulc, S.~Backhaus, and M.~Chertkov, ``Options for control of reactive power by distributed photovoltaic generators,'' \emph{Proceedings of the {IEEE}}, vol.~99, no.~6, pp. 1063--1073, June 2011.

\bibitem{Agalgaonkar_2014}
Y.~P. Agalgaonkar, B.~C. Pal, and R.~A. Jabr, ``Distribution voltage control considering the impact of {PV} generation on tap changers and autonomous regulators,'' \emph{{IEEE} Transactions on Power Systems}, vol.~29, no.~1, pp. 182--192, Jan. 2014.

\bibitem{Baran_1989_loss}
M.~Baran and F.~Wu, ``Network reconfiguration in distribution systems for loss reduction and load balancing,'' \emph{IEEE Transactions on Power Delivery}, vol.~4, no.~2, pp. 1401--1407, Apr. 1989.

\bibitem{IEEE_2018}
IEEE, ``Ieee standard for interconnection and interoperability of distributed energy resources with associated electric power systems interfaces,'' \emph{IEEE Std 1547-2018 (Revision of IEEE Std 1547-2003)}, vol.~2, no.~1, pp. 1--138, April 2018.

\bibitem{Naumann_2015}
M.~Naumann, R.~C. Karl, C.~N. Truong, A.~Jossen, and H.~C. Hesse, ``Lithium-ion battery cost analysis in pv-household application,'' \emph{Energy Procedia}, vol.~73, pp. 37--47, jun 2015, 9th International Renewable Energy Storage Conference, IRES 2015.

\bibitem{2015_Schmidt}
J.~P. Schmidt, A.~Weber, and E.~Ivers-Tiffée, ``A novel and fast method of characterizing the self-discharge behavior of lithium-ion cells using a pulse-measurement technique,'' \emph{Journal of Power Sources}, vol. 274, pp. 1231--1238, jan 2015.

\bibitem{monte_1949}
\BIBentryALTinterwordspacing
N.~Metropolis and S.~Ulam, ``The monte carlo method,'' \emph{Journal of the American Statistical Association}, vol.~44, no. 247, pp. 335--341, 1949. [Online]. Available: \url{http://www.jstor.org/stable/2280232}
\BIBentrySTDinterwordspacing

\bibitem{Sutton_2018}
R.~Sutton and A.~Barto, \emph{Reinforcement Learning, second edition: An Introduction}, ser. Adaptive Computation and Machine Learning series.\hskip 1em plus 0.5em minus 0.4em\relax MIT Press, 2018.

\bibitem{Marra_2014}
F.~Marra, G.~Yang, C.~Traeholt, J.~Ostergaard, and E.~Larsen, ``A decentralized storage strategy for residential feeders with photovoltaics,'' \emph{{IEEE} Transactions on Smart Grid}, vol.~5, no.~2, pp. 974--981, March 2014.

\bibitem{SMA_SmartHome_2023}
\BIBentryALTinterwordspacing
{SMA Solar Technology}, \emph{SMA SMART HOME: Battery Charging Management with Time-of-Use Energy Tariffs}, Niestetal, Deutschland, 2023. [Online]. Available: \url{www.sma.de}
\BIBentrySTDinterwordspacing

\bibitem{MendezQuezada_2006}
V.~Quezada, J.~Abbad, and T.~Roman, ``Assessment of energy distribution losses for increasing penetration of distributed generation,'' \emph{IEEE Transactions on Power Systems}, vol.~21, no.~2, pp. 533--540, 2006.

\bibitem{Carvalho_2008}
P.~Carvalho, P.~Correia, and L.~Ferreira, ``Distributed reactive power generation control for voltage rise mitigation in distribution networks,'' \emph{{IEEE} Transactions on Power Systems}, vol.~23, no.~2, pp. 766--772, May 2008.

\bibitem{EnWG2005}
\BIBentryALTinterwordspacing
G.~F.~M. of~Justice~(BMJ), ``Act on the supply of electricity and gas (energy industry act - enwg),'' \emph{BMJ}, vol. 272, no.~1, pp. 1--39, Oct. 2005, date of issue: 07.07.2005. Full citation: "Energy Industry Act of 7 July 2005 (BGBl. I p. 1970; 3621), as last amended by Article 9 of the Act of 26 July 2023 (BGBl. 2023 I No. 202)". Status: Last amended by Art. 9 G v. 12.10.2023 I No. 272. [Online]. Available: \url{www.gesetze-im-internet.de}
\BIBentrySTDinterwordspacing

\bibitem{Brown_2008}
R.~E. Brown, ``Impact of smart grid on distribution system design,'' in \emph{2008 IEEE Power and Energy Society General Meeting - Conversion and Delivery of Electrical Energy in the 21st Century}.\hskip 1em plus 0.5em minus 0.4em\relax IEEE, Jul. 2008.

\bibitem{Dorfler_2019}
F.~Dorfler, S.~Bolognani, J.~W. Simpson-Porco, and S.~Grammatico, ``Distributed control and optimization for autonomous power grids,'' in \emph{2019 18th European Control Conference (ECC)}.\hskip 1em plus 0.5em minus 0.4em\relax IEEE, Jun. 2019.

\bibitem{Mueller_2023}
F.~Mueller, S.~de~Jongh, X.~Mu, M.~Suriyah, and T.~Leibfried, ``Sector-coupled distribution grid analysis for centralized and decentralized energy optimization,'' in \emph{2023 58th International Universities Power Engineering Conference (UPEC)}.\hskip 1em plus 0.5em minus 0.4em\relax IEEE, Aug. 2023.

\bibitem{2020_wolfle}
\BIBentryALTinterwordspacing
D.~W\"{o}lfle, A.~Vishwanath, and H.~Schmeck, ``A guide for the design of benchmark environments for building energy optimization,'' in \emph{Proceedings of the 7th ACM International Conference on Systems for Energy-Efficient Buildings, Cities, and Transportation}, ser. BuildSys '20.\hskip 1em plus 0.5em minus 0.4em\relax New York, NY, USA: Association for Computing Machinery, 2020, p. 220–229. [Online]. Available: \url{https://doi.org/10.1145/3408308.3427614}
\BIBentrySTDinterwordspacing

\bibitem{Beiranvand2017}
V.~Beiranvand, W.~Hare, and Y.~Lucet, ``Best practices for comparing optimization algorithms,'' \emph{Optimization and Engineering}, vol.~18, no.~4, pp. 815--848, 2017.

\bibitem{Braun_2009}
M.~Braun, T.~Stetz, T.~Reimann, B.~Valov, and G.~Arnold, ``Optimal reactive power supply in distribution networks - technological and economic assessment for pv systems,'' in \emph{Proceedings of the 24th European Photovoltaic Solar Energy Conference (EU PVSEC 2009)}, Hamburg, Germany, Sep. 2009.

\end{thebibliography}
	
	% Author Biographies
	% Author(s): Gökhan Demirel
% Author Affiliation:
% Institute for Automation and Applied Informatics (IAI),
% Karlsruhe Institute of Technology (KIT),
% Eggenstein-Leopoldshafen, Germany
%!TEX root = ./Impact_and_Integration_of_Mini_Photovoltaic_Systems_on_Electric_Power_Distribution_Grids.tex
% IEEEbio_authors.tex
\vspace{-15mm}
\begin{IEEEbiography}[{\includegraphics[width=1.00in,height=1.25in,clip,keepaspectratio]{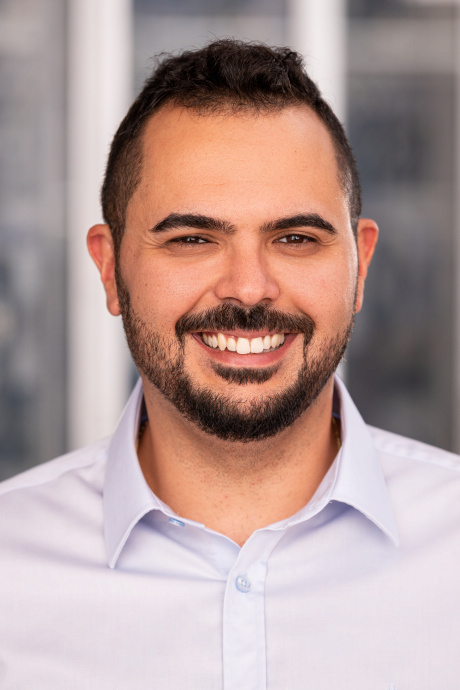}}]{Gökhan Demirel}
received the B. Sc. and M. Sc. degrees in electrical engineering and information technology from the Karlsruhe Institute of Technology (KIT), Karlsruhe, Germany, in 2017 and 2021.
He is currently pursuing the Ph.D. degree in electrical engineering with Karlsruhe Institute of Technology, Karlsruhe, Germany. 
His research interests include optimal control, smart grid, reinforcement learning, and their applications.
\end{IEEEbiography}
\vspace{-15mm}
\begin{IEEEbiography}[{\includegraphics[width=1.00in,height=1.25in,clip,keepaspectratio]{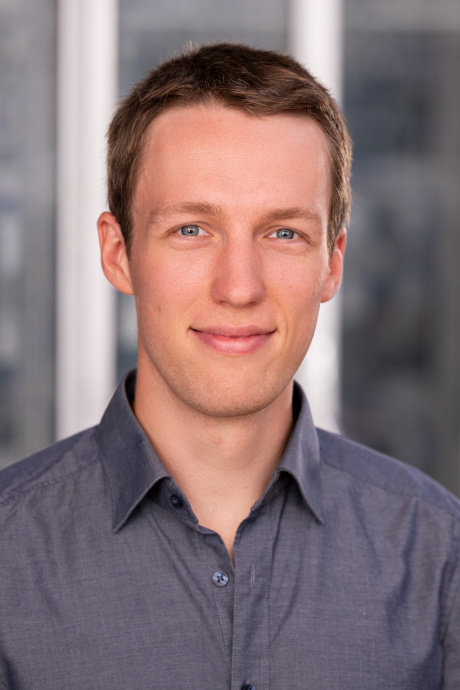}}]{Simon Grafenhorst} 
received the B. Sc. and M. Sc. degrees in computer science from the Karlsruhe Institute of Technology (KIT), Karlsruhe, Germany, in 2016 and 2020.
He is currently pursuing the Ph.D. degree in computer science with Karlsruhe Institute of Technology, Karlsruhe, Germany.
His research interests include the optimization of multi-energy systems in the distribution grid. 
\end{IEEEbiography}
\vspace{-15mm}
\begin{IEEEbiography}[{\includegraphics[width=1.00in,height=1.25in,clip,keepaspectratio]{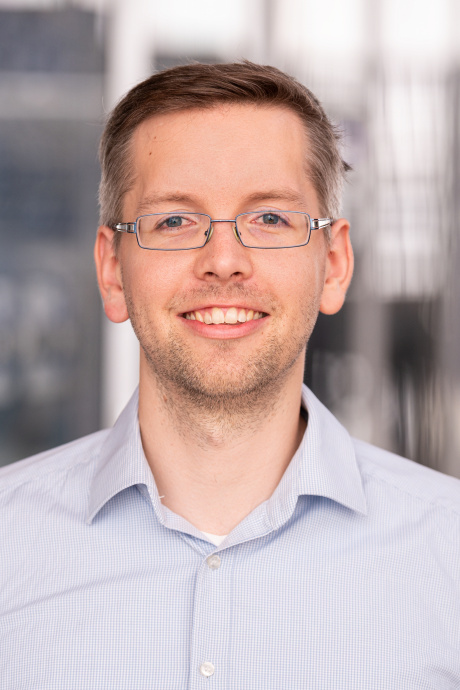}}]{Kevin Förderer}
received the M.Sc. degree in economathematics from the Karlsruhe Institute of Technology (KIT), Karlsruhe, Germany, in 2016 and the Dr.-Ing. from KIT, Germany, in 2021.
He is currently the Head of Group for the research group IT Methods and Components for Energy Systems (IT4ES), Institute for Automation and Applied Informatics, KIT, Karlsruhe, Germany. 
His research interests include demand side management and flexibility modeling in smart grids using AI as surrogate models.
\end{IEEEbiography}
\vspace{-15mm}
\begin{IEEEbiography}[{\includegraphics[width=1.00in,height=1.25in,clip,keepaspectratio]{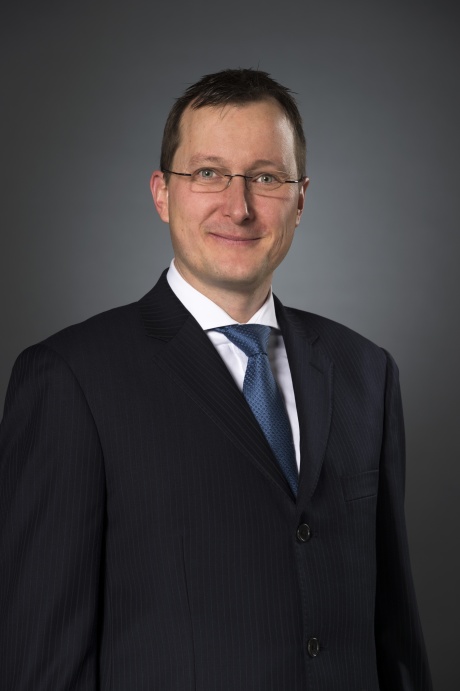}}]{Veit Hagenmeyer}
(Member, IEEE) received the Ph.D. degree from the Laboratoire des Signaux et Syst\`{e}mes (L2S), C.N.R.S.-Sup\'{e}lec,Universit\'{e} Paris-Sud, Bures-sur-Yvette, France, in 2002. 
He is currently the Professor of Energy Informatics with the Faculty of Informatics, and the Director of the Institute for Automation and Applied Informatics, Karlsruhe Institute of Technology, Karlsruhe, Germany. 
His research interests include modeling, optimization and control of sector-integrated energy systems, machine-learning based forecasting of uncertain demand and production in energy systems mainly driven by renewables, and integrated cyber-security of such systems.
\end{IEEEbiography}
\vspace{-10mm}
	
\end{document}